\documentclass[twocolumn]{aastex701}

\usepackage{threeparttable}
\usepackage{bm}
\usepackage{subcaption}
\usepackage{amsmath}
\usepackage{graphicx}
\usepackage{booktabs}
\usepackage{threeparttable}
\usepackage{siunitx}
\usepackage{multirow}
\usepackage{tabularx}
\usepackage{caption}
\usepackage{float}
\usepackage{lineno}
\usepackage{soul}    
\usepackage{xcolor}   
\sethlcolor{yellow}
\usepackage{ulem}

\begin{document}

\title{Hard X-ray Emission in AU Mic Flares: A Minor Contributor to Planetary Atmospheric Escape}

\author[orcid=0009-0006-2331-651X]{Yifan Hu}
\affiliation{Cahill Center for Astrophysics, California Institute of Technology, 1216 East California Boulevard, Pasadena, CA 91125, USA}
\affiliation{Department of Physics, Imperial College London, Exhibition Road, London SW7 2AZ, UK}
\email[show]{yh3123@ic.ac.uk}  

\author[orcid=0000-0002-8147-2602]{Murray Brightman} 
\affiliation{Cahill Center for Astrophysics, California Institute of Technology, 1216 East California Boulevard, Pasadena, CA 91125, USA}
\email{mbright@caltech.edu}

\author[orcid=0009-0001-8144-2526]{Fabio Favata}
\affiliation{Department of Physics, Imperial College London, Exhibition Road, London SW7 2AZ, UK}
\affiliation{INAF—Osservatorio Astronomico di Palermo, Piazza del Parlamento, 1, 90134 Palermo, Italy}
\email{f.favata@imperial.ac.uk}

\author[]{Haiwu Pan}
\affiliation{National Astronomical Observatories, Chinese Academy of Sciences, Beijing 100101, People’s Republic of China}
\email{}

\author[orcid=0000-0002-1984-2932]{Brian Grefenstette}
\affiliation{Cahill Center for Astrophysics, California Institute of Technology, 1216 East California Boulevard, Pasadena, CA 91125, USA}
\email{}

\author[orcid=0000-0002-4226-8959]{Fiona A. Harrison}
\affiliation{Cahill Center for Astrophysics, California Institute of Technology, 1216 East California Boulevard, Pasadena, CA 91125, USA}
\email{}

\author[orcid=0000-0003-2686-9241]{Daniel Stern}
\affiliation{Jet Propulsion Laboratory, California Institute of Technology, 4800 Oak Grove Drive, Pasadena, CA 91109, USA}
\email{}

\author[]{Weimin Yuan}
\affiliation{National Astronomical Observatories, Chinese Academy of Sciences, Beijing 100101, People’s Republic of China}

\affiliation{School of Astronomy and Space Science, University of Chinese Academy of Sciences, Chinese Academy of Sciences, Beijing 100049, People’s Republic of China}
\email{}

\author[orcid=0000-0002-4263-2562]{Yuk L. Yung}
\affiliation{Division of Geological and Planetary Science, California Institute of Technology, Pasadena, CA 91125, USA}
\email{}

\author[orcid=0000-0002-7791-3671]{Xiurui Zhao}
\affiliation{Cahill Center for Astrophysics, California Institute of Technology, 1216 East California Boulevard, Pasadena, CA 91125, USA}
\email{}

\begin{abstract}
Stellar flares are potent drivers of atmospheric evolution on orbiting exoplanets, primarily through extreme ultraviolet (EUV) and soft X-ray (XUV) irradiation. However, the contribution of hard X-rays (HXR; 3--20 keV)—which penetrate deeper into planetary atmospheres—to mass loss and particle acceleration has remained poorly understood. To quantify the HXR share of the total radiative budget, we conducted quasi-simultaneous observations of the active M-dwarf AU Mic using NuSTAR, Swift, and the Einstein Probe. Our analysis detected two major flares, and we performed an empirical check by deriving a quiescent-phase soft X-ray (SXR; 0.3--3 keV)–HXR relation and then applying it to the flares. By combining this with the quiescent coronal SXR–EUV relations conversion of \citet{sanzforcada2011}, we computed the total high-energy flux (EUV + SXR + HXR) and assessed the relative role of HXR in atmospheric escape. We find that HXR accounts for only a few percent of the total radiative energy budget during both quiescent and flaring states. While a high-energy spectral tail is detected in the second flare, time-resolved spectroscopy reveals a dominant chromospheric-evaporation signature, indicating that the flare energetics are primarily thermal.

\end{abstract}

\keywords{\uat{Stellar activity}{1580} --- \uat{Stellar flares}{1603} --- \uat{X-ray transient sources}{1852} --- \uat{Star-planet interactions}{2177}}


\section{Introduction} 

Stellar flares are intense, transient bursts of electromagnetic radiation triggered by sudden releases of magnetic energy in stellar atmospheres (See review: \citealt{Favata_2003,Gudel:2004bz,Benz2017,kowalski2024stellarflares}; See reference: \citealt{Sturrock1966,Vaiana1981}). These events, observed across all wavelength bands, from X-rays to radio, originate from magnetic reconnection processes driven by convective turbulence and shear in rotating stellar interiors (See review: \citealt{Favata_2003,Gudel:2004bz,Benz2017,kowalski2024stellarflares}; See reference: \citealt{Kopp1976}). In well-established models developed from solar observations, a significant fraction ($\approx$ 40\%) of the released magnetic energy is channeled into accelerated electrons \citep{Aschwanden_2016}. These nonthermal electrons produce hard X-ray (HXR) emission via bremsstrahlung as they collide with dense chromospheric material \citep{brown1971deduction}, while simultaneously heating the local plasma to temperatures exceeding 10 million Kelvin (MK). The resulting overpressure drives chromospheric evaporation into coronal loops, leading to soft X-ray (SXR) emission from a hot, thermal plasma. Together, these processes give rise to the complex, multi-wavelength signatures of stellar flares \citep{Gudel:2004bz}. While much progress has been made in understanding solar flares, key questions remain about how these energy release mechanisms scale to young, active stars. In particular, quantifying the role of HXR in stellar flares is essential, not only for constraining particle acceleration and flare energetics, but also because high-energy radiation can strongly impact the upper atmospheres of orbiting exoplanets \citep{Segura2010}. While extreme ultraviolet (EUV) and SXR, together referred to as XUV, are typically assumed to dominate the energy deposition, HXR emission may contribute via deeper atmospheric penetration and heating, especially during strong flares, a contribution that has previously remained observationally unconstrained. This connection becomes especially important when considering the long-term atmospheric retention and habitability of planets around magnetically active hosts.

M-dwarfs are low-mass ($0.1$--$0.6 \,M_\odot$), cool ($T_{\rm eff} \lesssim 3900$\ K) stars and constitute the most numerous stellar population in the Galaxy, making up more than 70\% of all stars \citep{Henry2024}. They exhibit strong, long-lived magnetic activity due to their deep convective envelopes and inefficient angular momentum loss \citep{West_2008,Newton_2016}. As a result, flares on M dwarfs are not only more frequent but also significantly more energetic than on their solar counterparts, sometimes releasing over $10^{34}$ erg in a single event, whereas even the largest solar flares rarely exceed $10^{32}$ erg. \citep{kane2005multispacecraft,Osten_2015,Davenport_2016}. These extreme conditions provide a natural laboratory for probing magnetic reconnection and particle acceleration physics beyond the solar regime \citep{Gudel:2004bz,Benz2017}. At the same time, the high flare activity of M dwarfs is particularly relevant to exoplanetary systems, as most of their habitable zones lie within $\approx$ 0.1--0.4 AU, where planets are exposed to intense stellar radiation and particle fluxes that can drive atmospheric escape and alter long-term climate stability \citep{Segura2010}.

The young ($\approx$ 23\,Myr), active M1Ve star AU Microscopii (AU Mic) exemplifies these characteristics \citep{plavchan2020nature}. AU Mic has a mass of $\approx$ 0.50 $M_{\odot}$ and a radius of $\approx$ 0.75 $R_{\odot}$. Located just 9.72\,pc away, AU Mic hosts three confirmed transiting exoplanets, AU Mic b, c, and d \citep{plavchan2020nature,wittrock2023validatingaumicroscopiid}. The classical habitable zone of AU Mic is estimated to span $\approx$ 0.31–0.60 astronomical units (AU), but all three planets orbit well inside this range: planet b at $\approx$ 0.066 AU, c at $\approx$ 0.11 AU, and d even closer, placing them far inside to the habitable zone and subjecting them to intense high-energy irradiation \citep{plavchan2020nature,Kane_2021}. AU Mic's close distance minimizes interstellar absorption, allowing high-energy photons emitted during flares, particularly in the X-ray regime, to reach Earth-based and orbital observatories with minimal attenuation. This is critical for SXR studies, since flux from more distant M‑dwarfs is often suppressed by the interstellar medium \citep{Gudel:2004bz}. Also, for AU Mic, The Nuclear Spectroscopic Telescope Array (NuSTAR) \citep{harrison2013} can directly observe HXR component, enabling us to quantify the nonthermal electron contribution to flare energy budgets for the first time in an exoplanet-hosting M dwarf. Such quantification is crucial for constraining atmospheric mass-loss models, particularly in the context of young, active stars where photoevaporation and flare-driven escape may be most pronounced \citep{Airapetian_2019}.

Previous studies have made significant progress in modeling and characterizing the flare-driven high-energy radiation environment of M dwarfs and its impact on planetary atmospheres. \cite{amaral2025impactstellarflaresatmospheric} combined time-dependent flare frequency distributions with stellar evolution and escape models to simulate mass loss from AU Mic’s planets, finding that quiescent XUV emission alone is sufficient to strip atmospheres from close-in planets like AU Mic d, while flare-driven radiation becomes essential for atmospheric loss in habitable-zone orbits during the post-saturation phase. Observationally, \cite{Feinstein_2022} performed time-resolved HST/COS observations of AU Mic in the far-UV, detecting 13 flares and showing that flare-driven radiation can enhance atmospheric escape rates by up to six orders of magnitude. However, due to limited sensitivity at higher X-ray energies, these prior works do not measure or constrain the contribution of HXR or the energy carried by nonthermal electrons.

NuSTAR is the first focusing HXR telescope in orbit offering enhanced sensitivity above 10 keV compared to previous X-ray missions. Its improved coverage of the high-energy regime enables direct measurements of nonthermal bremsstrahlung emission during stellar flares and allows for the detection of hot plasma continua above 10~MK. These capabilities make NuSTAR a unique tool for probing flare energetics and assessing the role of HXR radiation in driving atmospheric escape from close-in exoplanets. When combined with observations from the Neil Gehrels Swift Observatory (Swift) \citep{Gehrels_2004} and the Einstein Probe (EP) \citep{yuan2022}, this multi-wavelength approach enables a quantitative assessment of the HXR contribution during flares. For the first time, we leverage overlapping observations by Swift, EP, and NuSTAR to empirically establish a SXR to HXR scaling relation in a stellar flare, enabling constraints on the hard X-ray contribution to planetary atmospheric escape. 

The paper is organized as follows: Section \ref{sec:obs} introduces the observations and outlines the process for data reduction and extraction. Section \ref{sec:analysis} describes the analysis of flare spectra, and the corresponding results are presented. Section \ref{sec:discussion} discusses of the results, and Section \ref{sec:summary} provides a summary of our study.

\section{Observations and Data Reduction} \label{sec:obs}

AU Mic was observed three times by X-ray instruments during this campaign: two SXR observations with Swift X‐Ray Telescope (XRT), one SXR observation with EP Follow-up X-Ray Telescope (FXT), and one HXR observation with NuSTAR. The details of these observations are given in Table \ref{tab:au_mic_obs}.

Our first monitoring epoch was a Swift–XRT Target of Opportunity (ToO) on UT 2025 June 30 (ToO ID 22793; PI: Y. Hu), comprising a total of 16 ks of XRT exposure (Photon-Counting mode), which was split into two sequences. The follow-up NuSTAR Director’s Discretionary Time (DDT) program (PI: Y. Hu) was executed on UT 2025 July 1, with a net exposure of 46 ks. Finally, an EP–FXT ToO observed AU Mic on UT 2025 July 2 (PI: H. Pan) in Full Frame Mode (FF)+thin-filter mode for 6 ks.

Because the count rates of NuSTAR were low, leaving many spectral bins with only a few photons, we performed spectral fitting by minimizing the Poisson‐based Cash statistic ($C_{\rm stat}$; \citealt{Cash1979}), which in XSPEC is implemented as the W‐statistic when a background spectrum is subtracted. Spectral fitting was carried out in XSPEC v12.14.1 \citep{arnaud1996} under HEASoft 6.34. 

\begin{deluxetable*}{c c c c c c c}[h]
\tabletypesize{\scriptsize}
\tablewidth{0pt}
\tablecaption{Summary of AU\,Mic X-ray observations\label{tab:au_mic_obs}}
\tablehead{
  \colhead{Observatory} &
  \colhead{Energy Range (keV)} &
  \colhead{Obs ID} &
  \colhead{Mode} &
  \colhead{Start Date \& Time (UTC)} &
  \colhead{End Date \& Time (UTC)} &
  \colhead{Exp.\ Time (ks)\tablenotemark{a}}
}
\startdata
Swift–XRT   &0.3--10.0  & 03400165001    &  PC\tablenotemark{b}           & 2025-06-30 14:38:26 & 2025-06-30 22:51:46 & 9.4 \\
Swift–XRT   &0.3--10.0  & 03400165002    &  PC           & 2025-07-01 00:09:44 & 2025-07-01 17:29:44 & 6.7 \\
NuSTAR     &3--79   & 91101320002  &  PC           & 2025-07-01 15:44:36 & 2025-07-02 15:51:23 & 46.5 \\
EP–FXT     &0.5--10.0   & 06800000712      &  FF\tablenotemark{c} + thin    & 2025-07-02 10:49:20 & 2025-07-02 13:15:30 &  6.0 \\
\enddata
\tablenotetext{a}{Total scheduled exposure time. Cleaned exposure time may be slightly shorter due to background filtering.}
\tablenotetext{b}{Photon Counting mode, providing event-by-event photon detection with high timing accuracy.}

\tablenotetext{c}{Einstein Probe Full Frame mode, with full CCD readout and moderate time resolution.}

\end{deluxetable*}

\subsection{Swift–XRT Data Reduction} \label{sec:Swift_data_reduction}

Swift–XRT observations were processed with XRT Data Analysis Software (XRTDAS) v3.6.1 and calibrated against CALDB20250609 using the \texttt{xrtpipeline} task. Level 2 event files were then filtered in \texttt{xselect} to define source and background regions: a $20''$ radius circle centered on AU Mic for source spectra, corresponding to $\approx 1.1$ times the Swift–XRT half-power diameter (HPD $\approx 18''$; \citealt{Gehrels_2004}), thus enclosing $\gtrsim 50\%$ of the source photons while minimizing background contamination, and an $80''$ radius circle located $50''$ away in a source‐free region for the background. The extracted source and background spectra were taken from \texttt{xrtmkarf} to generate ancillary response files (ARFs), while the appropriate response matrix files (RMFs) were retrieved from CALDB.  All spectra, ARFs, and RMFs were then grouped with \texttt{GRPPHA} with minimum 1 count for each bin for all spectra.

Light curves were produced by extracting source and background light curves from the same regions in \texttt{xselect}, then using exposure maps for vignetting and bad‐pixel corrections. Background subtraction and exposure correction were carried out with \texttt{lcmath}, yielding background‐subtracted, exposure‐corrected source light curves for all timing and flare‐identification analyses. With peak count rates of $\approx0.4\,$ct\,s$^{-1}$, below the 0.5 ct\,s$^{-1}$ pile‐up threshold, no pile‐up correction was necessary.

The two consecutive exposures (Table \ref{tab:au_mic_obs}) both provided sufficient counts for independent spectral fits, and slight shifts in the source position between the observations necessitated separate extraction regions; accordingly, the exposures were analyzed independently.

\subsection{EP-FXT Data Reduction}

EP-FXT employs nested Wolter‐I optics divided into two independent units (FXT‐A and FXT‐B). Each unit comprises three assemblies: an upper composite (mirror modules), a lower composite (detector assembly), and a supporting structure.

We processed the EP–FXT data using FXT Data Analysis Software (FXTDAS) v1.20 together with the FXT CALDB v1.20. Event files were calibrated via the standard pipeline task. Source spectra were extracted from a circular region of radius $13''$ centered on AU Mic, smaller than the nominal HPD ($\approx30''$) to reduce background and possible contamination from nearby sources while retaining high signal-to-noise ratio given the source brightness \citep{yuan2022}. Background spectra were extracted from a $40''$ radius circular region located $30''$ away from the source position in a source-free area, carefully avoiding any resolved sources. With a source count rate of $2\ \mathrm{ct\,s^{-1}}$, below the FXT pile‐up threshold, no pile‐up correction was applied, following the guidelines in the EP–FXT Users Guide v1.20 \citep{epfxt2025}.

EP–FXT level 2 data were reduced following the same procedure as described in Section \ref{sec:Swift_data_reduction}. In brief, after events were filtered, source and background regions defined, spectra were extracted, and ARFs and RMFs generated before grouping with {\tt GRPPHA}. Light curves were corrected using exposure maps and background‐subtracted via {\tt lcmath}. The ARFs and RMFs were generated by \texttt{fxtarfgen} and \texttt{fxtrmfgen} by FXTDAS, respectively.

\subsection{NuSTAR Data Reduction}

The NuSTAR payload comprises two co-aligned grazing-incidence optics modules sensitive over 3–79 keV \citep{harrison2013}. Each focal plane module (FPM-A and FPM-B) employs a $2\times2$ array of pixellated cadmium–zinc–telluride detectors, producing small cross-shaped gaps between detector quadrants.

We retrieved the calibrated event files for both FPM-A and FPM-B and processed them with NuSTAR Data Analysis Software (NUSTARDAS) v2.1.4 and CALDB 20250224 via the \texttt{nupipeline} task \citep{nustardas2021}. Source and background spectra were then extracted using \texttt{nuproducts}. For each module, the source spectrum was extracted from a circular region of radius $50''$ centered on the source; the background was extracted from a $120''$ radius circular region located $50''$ away from the source center, on the same detector quadrant, and chosen to be free of any source contamination, following the guidelines in the NuSTAR Data Analysis Quickstart Guide \citep{nustardas2021}. NuSTAR has an angular resolution of $18''$ (FWHM) and a half‐power diameter (HPD) of $58''$ \citep{harrison2013}, and the chosen extraction radius encloses the majority of the source counts while minimizing background contribution.  

NuSTAR level 3 data products were processed with the {\tt nuproducts} pipeline to extract both spectra and light curves directly from predefined source and background regions. In Section \ref{sec:flare_identification} and \ref{sec:flare_analysis}, after selecting the flare time intervals, we applied the same {\tt nuproducts} procedure, using identical regions for source and background, to extract the corresponding flare‐interval spectra and light curves. We used \texttt{lcurve} with multiple bin sizes to verify presence of the flare. This ensured full consistency between the integrated and time‐resolved spectral products.

\section{Data Analysis and Results} \label{sec:analysis}

\subsection{Energy Band Definitions}

Previous investigations of atmospheric escape from M‐dwarf exoplanets have focused almost exclusively on the XUV regime (0.01 to a few keV), because as photons in this band deposit their energy deep in the upper atmosphere of exoplanets, they heat and ionize the gas, driving powerful hydrodynamic escape \citep{lammer2003}.  However, HXR may deposit energy at greater depths or contribute to nonthermal escape channels, a possibility that remains largely unexplored. To quantify HXR’s potential role alongside the well‐studied XUV input, we subdivide our data into three instrument‐matched bands:

\paragraph{EUV Band (0.01--0.3 keV)}  
The long‐wavelength cutoff at 0.01 keV includes the hydrogen Lyman limit at 912 \AA ($\approx$ 0.014 keV), marking the transition from ultraviolet to EUV emission. The EUV band is characterized by strong spectral features, including the He II Ly $\alpha$ line at 304 \AA ($\approx$ 0.041 keV), strong coronal iron lines from Fe IX–Fe XIV ($\approx$ 170–284 \AA; $\approx$ 0.044–0.073 keV), and other transition-region lines. The high‐energy cutoff at 0.3 keV matches the low‐energy sensitivity limit of Swift–XRT \citep{burrows2005}, allowing us to bridge the EUV regime into the SXR band sampled by XRT. Given the limitations of current instrumentation, direct measurements of the EUV flux and related parameters are not feasible. Consequently, this study relies on the scaling relation between EUV and SXR to extrapolate the EUV properties.

\paragraph{SXR Band (0.3--3 keV)}  
Covering the 0.3–3 keV range, this band probes the bulk of thermal coronal plasma emission from AU Mic, including Fe XVII–Fe XXIV lines that dominate the 0.7–1.5 keV range, alongside significant lower-energy lines from O, Ne, and Mg \citep{Gudel:2004bz}. Swift–XRT’s optimized effective area and energy resolution fully sample this interval \citep{burrows2005}, while EP–FXT, though primarily designed for 0.3–10 keV, achieves sufficient sensitivity only above 0.5 keV \citep{yuan2022}. Combining both instruments, we obtain continuous coverage of key thermal diagnostics, enabling measurements of temperature distributions, emission measures, and time‐resolved flare evolution. This spectral interval is crucial for constraining coronal heating processes and flare energetics in M dwarfs like AU Mic \citep{Reale_2010}. 

\paragraph{HXR Band (3--20 keV)}  
Defined by NuSTAR’s reliable energy response from 3 keV to 20 keV, above which the combined instrumental and cosmic background dominates and the signal‐to‐noise ratio rapidly declines, this band captures both the hottest multi‑MK thermal bremsstrahlung emission and nonthermal emission from accelerated electrons \citep{Holman_2011}.

\medskip  
The adopted energy bands differ from conventions used in solar physics. In particular, the GOES 1--8 $\,\AA\,$ channel ($\approx$1.5--12\,keV) is often classified as part of the solar ''soft X-ray'' band; here, for clarity and to match instrument bandpasses, we define SXR as 0.3--3\,keV and HXR as 3--20\,keV.
These energy‐band definitions enable us to: (1) compare our observations directly with prior XUV‐based escape studies; (2) isolate the distinct heating regimes in AU Mic’s corona and flares; and (3) quantify the previously unmeasured HXR contribution to atmospheric energy deposition. Unless stated otherwise, all light curves, spectral fits, and fluxes in this work refer to these bands. 

\subsection{Flare Identification} \label{sec:flare_identification}

\begin{figure*}[ht] 
  \centering
  \includegraphics[width=\textwidth]{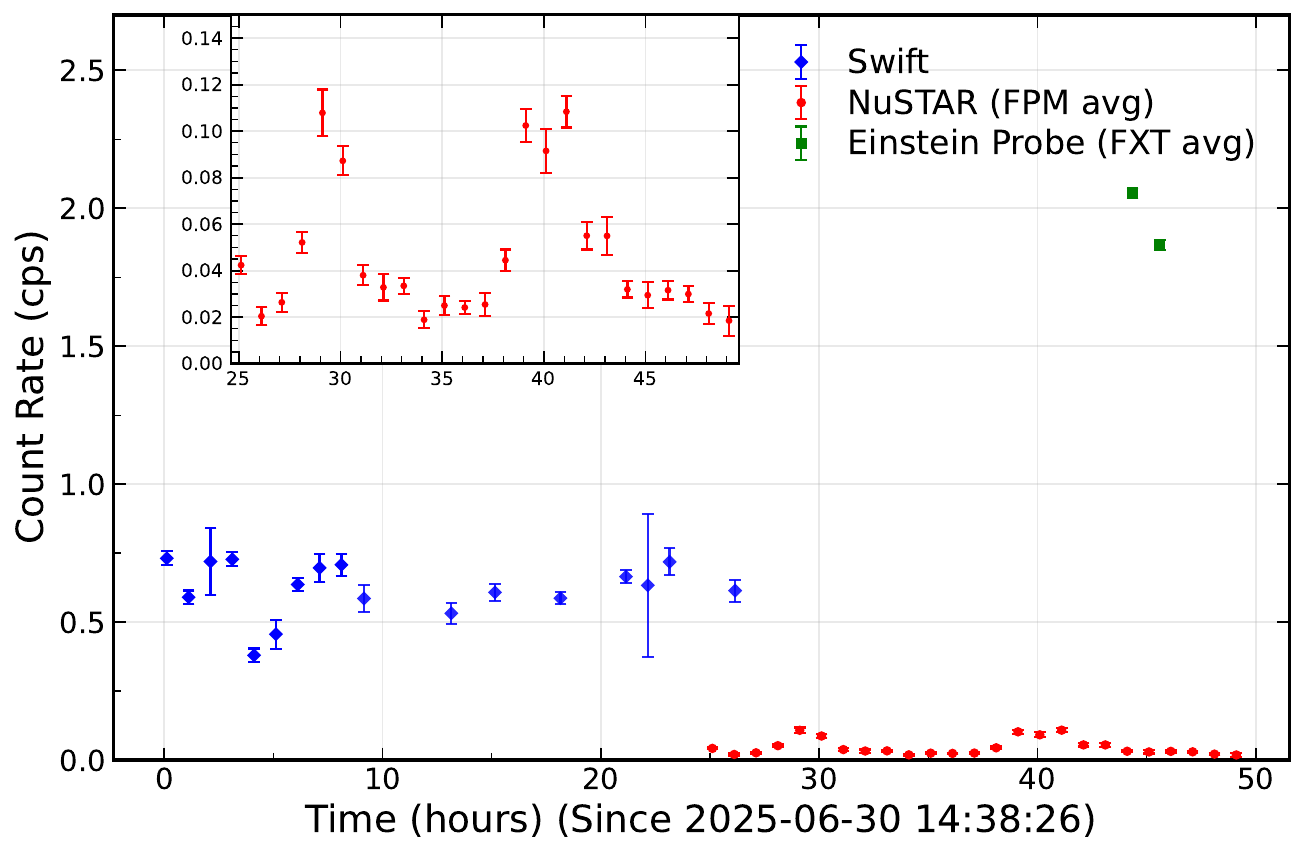}
  \caption{  
    Count rates from Swift, NuSTAR, and EP observations of AU Mic. The main panel shows background-subtracted Swift (0.3–10 keV), NuSTAR FPMA/B average (3–20 keV), and EP FXT-A/B average (0.5–10 keV) count rates, all binned at 3600 s. The 0–0.15 cps range that is expanded in the inset, which displays the NuSTAR data in greater detail. Error bars denote 1$\sigma$ uncertainties.
  }
  \label{fig:count_rates}
\end{figure*}

In this section, we present the detailed methodology employed in the identification of stellar flares within our observational datasets. The light curve of the three observations is shown in Figure \ref{fig:count_rates}. We applied iterative sigma clipping independently to the Swift and NuSTAR count-rate-based light curves. The primary purpose of this statistical approach was to determine the quiescent luminosity levels and their corresponding standard deviations ($\sigma$). For each dataset, the sigma-clipping was performed until convergence, ensuring the stability of the baseline count rate estimates.

While our default flare search was carried out on light curves binned at 3600 s, we also tested alternative bin sizes, including shorter bins of 450 s, 900 s, and 1800 s, as well as bins matched to the typical $\approx45$ min NuSTAR visibility window between Earth occultation. We found no qualitative change in the set of identified flares or in the subsequent analysis results. This demonstrates that our results are not sensitive to the adopted bin size.

\begin{figure}
    \centering
    \begin{subfigure}{\columnwidth}
        \centering
        \includegraphics[width=\columnwidth]{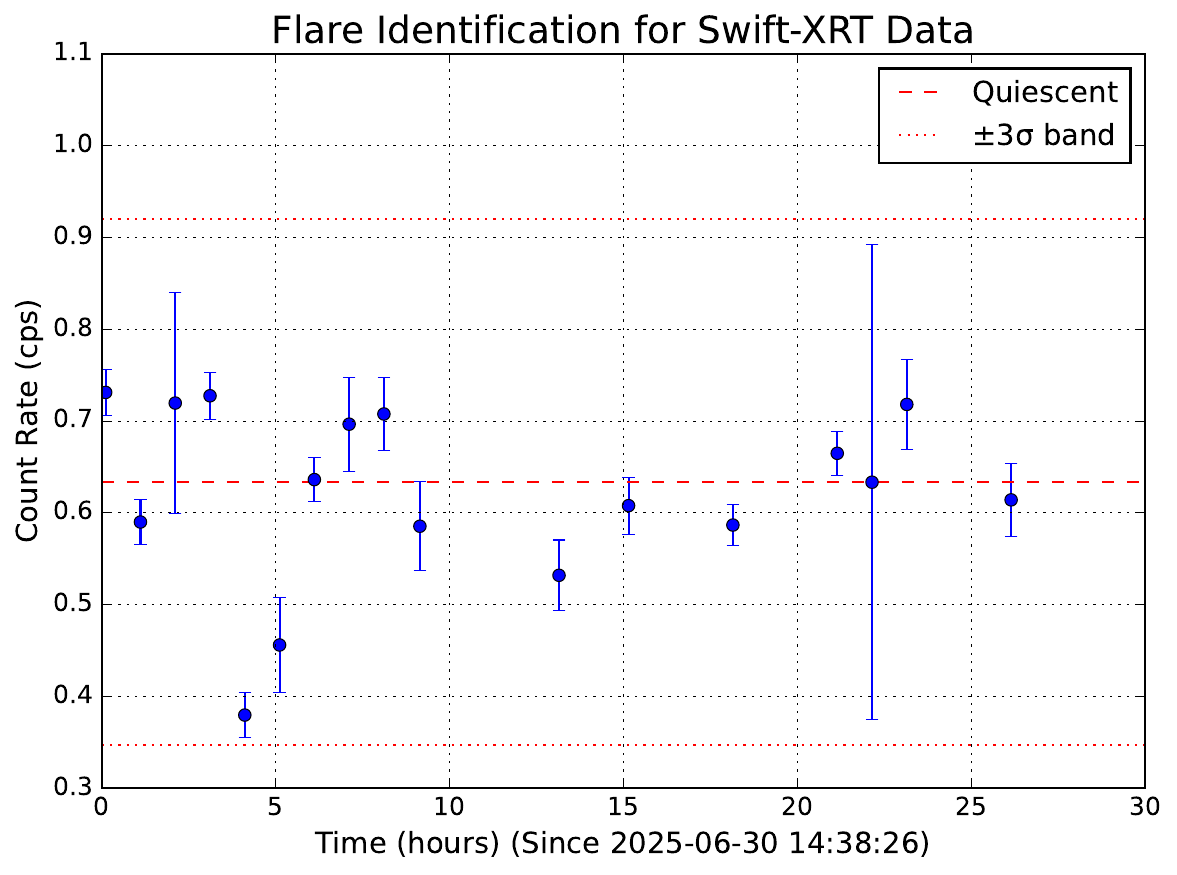}
    \end{subfigure}
    \vspace{0.5em}
    \begin{subfigure}{\columnwidth}
        \centering
        \includegraphics[width=\columnwidth]{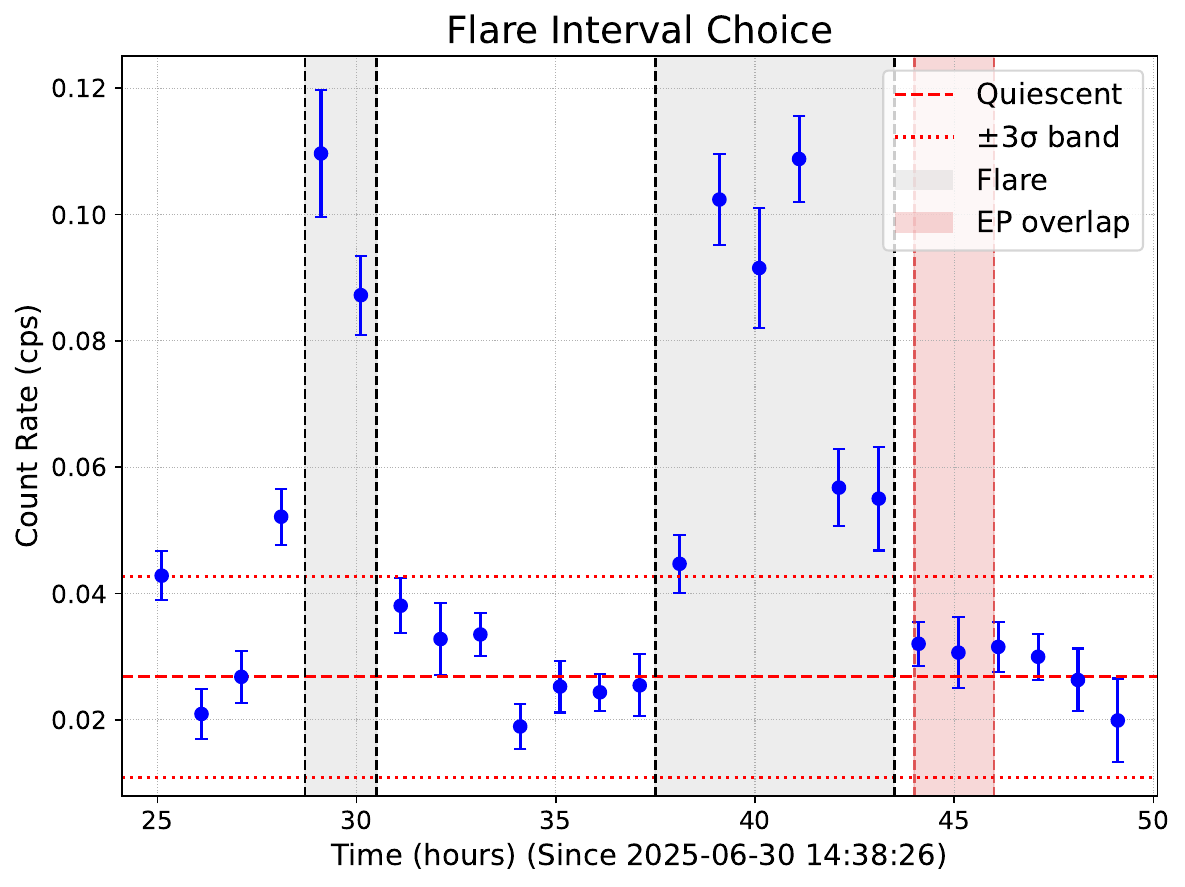}
    \end{subfigure}
    \caption{Top: Swift-XRT light curve, with the quiescent level (red dashed line) and ±3$\sigma$ bounds (dotted lines) estimated via one iteration of the 2$\sigma$-clipping method. Bottom: NuSTAR light curve with selected flare intervals shaded in gray. The red-shaded regions indicate intervals overlapping with the EP observations. Quiescent level and ±3$\sigma$ band are overlaid using the same sigma-clipping procedure, with seven iterations. Bin size = 3600 s.}
    \label{fig:flares_identification}
\end{figure}

The criterion for flare detection was then set as count rate exceeding the determined quiescent count rate plus three times the sigma value after clipping, which is the standard in \cite{Poyatos_2025}. Upon identifying potential flare candidates through this statistical threshold, we further visually inspected the corresponding sections of the light curves (bin size=200 s) to confirm the nature of these candidates as genuine flare events rather than statistical fluctuations or instrumental artifacts. This rigorous combined statistical and visual approach to flare identification is demonstrated explicitly in Figure \ref{fig:flares_identification}.

Based on these detection criteria, Swift did not capture any complete flare events. Even though partial flare stages may have been recorded, the available data do not conclusively support the interpretation of these events as complete flares. To avoid potential over-interpretation and ensure uniform analysis, we conservatively interpret the Swift data in the subsequent analysis to primarily only derive quiescent source properties.

\subsection{Spectral Analysis and Combined Fitting for Quiescent Spectrum}

\begin{table*}[htb]
\centering
\begin{threeparttable}
\caption{Quiescent spectral‐fit parameters for AU Mic}
\label{tab:specfit}
\begin{tabular*}{\textwidth}{@{\extracolsep{\fill}} llcccccc @{}}
\hline
Component   & Parameter         & Unit                                           & NuSTAR & EP–FXT & Swift Seq 1 & Swift Seq 2 \\
\hline
TBabs       & $N_{\rm H}$\tnote{a}       & $10^{20}\,\mathrm{cm^{-2}}$                    & 3.9\tnote{k}      & 3.9\tnote{k}       & $5.3_{-2.0}^{+2.5}$        & $2.6_{-1.9}^{+4.4}$        \\

APEC$_1$    & $T$\tnote{b}              & MK                                            & $19.9_{-2.2}^{+2.4}$    & $11.1_{-0.2}^{+0.2}$  & $16.1_{-1.4}^{+1.3}$              & $13.0_{-2.3}^{+2.3}$        \\
            & $Z$\tnote{c}         & —                               & $0.18_{-0.13}^{+0.19}$     & $0.10_{-0.01}^{+0.01}$ & $0.10_{-0.04}^{+0.05}$              & $0.06_{-0.02}^{+0.04}$        \\
            & $EM_1$\tnote{d}            & $10^{51}\,\mathrm{cm^{-3}}$                    & $5.5_{-1.1}^{+1.7}$  &
            $24.2_{-1.3}^{+1.3}$    & $17.6_{-2.2}^{+3.1}$             & $14.7_{-6.7}^{+9.7}$        \\

APEC$_2$    & $T$            & MK                                            & —      &$2.85_{-0.09}^{+0.09}$      & $3.74_{-0.57}^{+1.10}$        &  $6.13_{-3.44}^{+2.37}$        \\
            & $Z$\tnote{e}         & —                                 & —      & $0.10_{-0.01}^{+0.01}$      & $0.10_{-0.04}^{+0.05}$         & $0.06_{-0.02}^{+0.04}$         \\
            & $EM_2$           & $10^{51}\,\mathrm{cm^{-3}}$                    & —      & $38.0_{-2.2}^{+2.4}$      & $15.0_{-3.7}^{+10.5}$        &$12.0_{-3.5}^{+7.0}$        \\

power-law    & $\Gamma$\tnote{f}      & —                                              & 2 \tnote{l}        & —      & —        & —        \\
            & Norm\tnote{g}              & $10^{-4}$     & $0.5_{-0.3}^{+0.3}$      & —      & —        & —        \\

\hline
           & $N_{\rm dof}$\tnote{h}
            &—& 285 &672 & 327 & 257 \\
            &$C_{\rm stat}$\tnote{i}
            &—& 270.4 &776.9 & 321.9 & 261.6\\
            & Band\tnote{j}     & keV      & 3--20      & 0.5--3      & 0.3--3        & 0.3--3        \\
            & Average flux     & $10^{-12}\,\mathrm{erg\,cm^{-2}\,s^{-1}}$      & $0.7_{-0.1}^{+0.1}$       & $17.3_{-0.6}^{+0.6}$     & $13.0_{-0.5}^{+0.4}$\tnote{m}        & $12.3_{-0.4}^{+0.4}$\tnote{m}        \\
            & Power-law flux     & $10^{-12}\,\mathrm{erg\,cm^{-2}\,s^{-1}}$      & $0.2_{-0.1}^{+0.1}$       & —     & —       & —        \\
            & Average rate       & count\,s$^{-1}$                              & \begin{tabular}[t]{@{}l@{}}
                    0.011 (FPM‑A)\\
                    0.010 (FPM‑B)
                \end{tabular}       & \begin{tabular}[t]{@{}l@{}}
                    1.741 (FXT‑A)\\
                    2.223 (FXT‑B)
                \end{tabular} 

      & 0.404        & 0.385       \\
\hline
\end{tabular*}

\begin{tablenotes}[flushleft]
\item[] $\mathbf{Notes.}$ Uncertainties indicate the 90\% confidence interval. 
\item[a] $N_{\rm H}$: hydrogen column density.
\item[b] $T$: plasma temperature.
\item[c] Abundance relative to solar.
\item[d] EM: emission measure, adopting a distance of 9.72 pc \citep{plavchan2020nature}.
\item[e] Abundance fixed to the value of APEC$_2$.
\item[f] $\Gamma$: power‐law photon index.
\item[g] Power‐law normalization at 1 keV in units of photons keV$^{-1}$ cm$^{-2}$ s$^{-1}$.
\item[h] $N_{\rm dof}$: the number of independent spectral bins minus the number of free model parameters.
\item[i] $C_{\rm stat}$: Poisson‐based maximum‐likelihood statistic.
\item[j] Band for the observed flux (including Average flux and Power-law flux) and count rate.
\item[k] For NuSTAR and EP, $N_{\rm H}$ is fixed to the mean from the two Swift–XRT observations.
\item[l] Because the power-law component is unconstrained in the fit, we fix its photon index to 2, a typical value for nonthermal emission in stellar flares \citep{Benz2017}.
\item[m] Value derived from each model fit with fixed absorption $N_{\rm H}=3.9\times10^{20}\,$cm$^{-2}$ for consistency.
\end{tablenotes}
\end{threeparttable}
\end{table*}

We performed spectral fitting for quiescent periods of observations obtained by Swift–XRT, EP–FXT, and NuSTAR using XSPEC v12.14.1 \citep{arnaud1996}, adopting the $C_{\rm stat}$ appropriate for low-count data. For the two SXR telescopes, Swift–XRT and EP–FXT, spectral analysis was confined to energy ranges of 0.3–10 keV and 0.5–10 keV, respectively, based on instrument sensitivity. Prior to spectral fitting, we confirmed that in each selected energy band the source count rate, measured from the source extraction region, exceeded that from the background extraction region. The adopted model for both SXR spectra consists of an absorbed two-temperature collisionally ionized thermal plasma, represented as \texttt{tbabs}$\times$(\texttt{apec$_1$}+\texttt{apec$_2$}). Here, \texttt{tbabs} accounts for interstellar absorption with hydrogen column densities derived using the standard interstellar medium absorption model from \cite{Wilms_2000}, while each \texttt{apec} component \citep{Smith_2001} describes plasma emission characterized by temperature ($kT$), abundance relative to solar \citep{ANDERS1989197}, and normalization. Following spectral fitting, fluxes were computed over the energy ranges 0.3–3 keV (Swift–XRT) and 0.5–3 keV (EP–FXT), along with the corresponding instrument-specific count rates, as summarized in Table \ref{tab:specfit}.

For the NuSTAR observation, the emission was modeled using a single-temperature absorbed thermal plasma model because the NuSTAR band cannot constrain the lower temperature APEC. This model alone provides a statistically adequate fit capturing the continuum and prominent lines below 20 keV (The fit yields $C_{\rm stat}$ = 277.8 for $N_{\rm dof}$ = 286).

To investigate potential high-energy emission, we added a \texttt{power-law} component representing bremsstrahlung from nonthermal, accelerated electrons, and alternatively tested a thermal bremsstrahlung model \texttt{bremss} component to model high temperature thermal emission \citep{Osten_2007}. A narrow \texttt{Gaussian} line near 6.4–6.8 keV was also included to account for potential Fe K$\alpha$ fluorescence \citep{Testa2008}. While the Gaussian component did not yield a statistically improvement (The fit yields $C_{\rm stat}$ = 277.8 for $N_{\rm dof}$ = 285) and its normalization remained unconstrained, the inclusion of the \texttt{power-law} led to a notable decrease in the $C_{\rm stat}$ ($\Delta C_{\rm stat} = 7.4$), supporting its inclusion. However, due to the relatively low signal-to-noise ratio at higher energies, the photon index could not be well constrained and was therefore fixed at 2, a typical value observed in stellar flares \citep{Osten_2007}. Also, the \texttt{bremss} model yielded a comparable decrease in $C_{\rm stat}$ ($\Delta C_{\rm stat} = 7.4$) to that of the \texttt{power-law} component (the detailed parameters for the \texttt{bremss} component in quiescent period are listed in Table~\ref{tab:bremss}). However, during the quiescent period, the presence of a high temperature thermal plasma is physically implausible. As such, the final adopted model for the NuSTAR spectrum was \texttt{tbabs*(apec + power-law)}, which provides a better statistical fit without introducing unconstrained or physically unrealistic parameters.

\begin{figure}[htbp]
    \centering
    \includegraphics[width=\columnwidth]{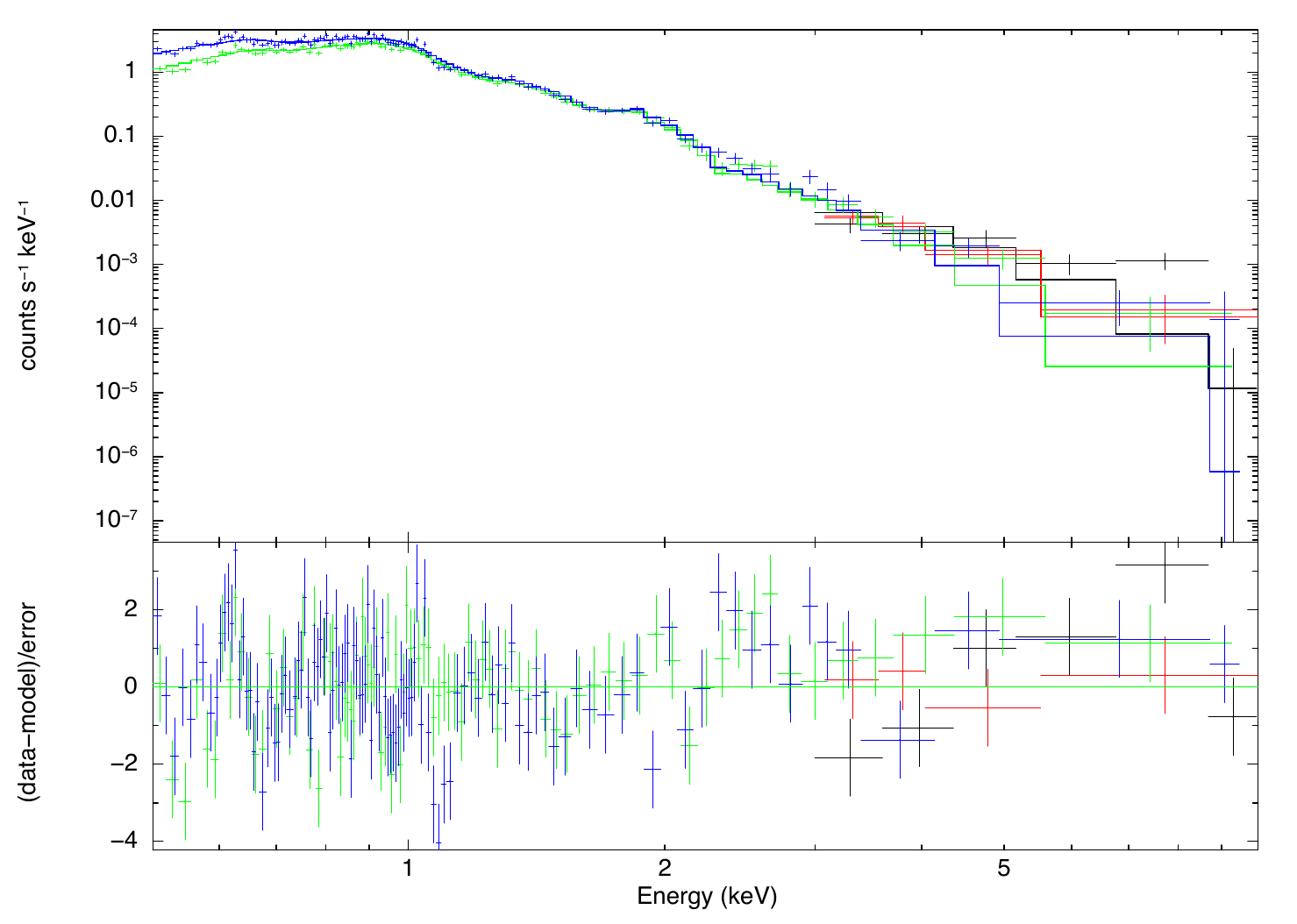}
    \caption{
        Joint spectral fitting results. Black points: FPM-A data;
        red points: FPM-B data;
        green points: FXT-A data;
        blue points: FXT-B data.
        The top panel shows the observed spectra and the best-fit folded model,
        while the bottom panel displays the residuals in units of $\sigma$.
    }
    \label{fig:joint_fit}
\end{figure}

From the spectral fitting of quiescent periods, we can calculate the energy conversion factor (ECF) by dividing the observed flux by the corresponding count rate, thus converting count-rate-based light curves into flux or luminosity plots. To accurately compare the SXR and HXR data, we must consider instrumental differences arising from calibration, effective area, and response functions. Therefore, joint spectral fitting is required to determine instrument-specific scaling factors. However, due to the limited overlap between the Swift–XRT and NuSTAR observations, we performed joint spectral fitting only for the simultaneous NuSTAR and EP–FXT data, shown in Figure \ref{fig:joint_fit}. We used NuSTAR FPM-A as the reference detector. Through this approach, we derived scaling factors for FPM-B, FXT-A, and FXT-B relative to FPM-A.

Due to the non-overlapping observational windows and significant fluctuations in the Swift–XRT light curves, we could not confirm whether the Swift data captured flare events, while EP–FXT did not. Consequently, joint fitting of these two instruments was not feasible. Instead, we computed Swift–XRT scaling factors as the mean of the FXT-A and FXT-B scaling factors. Detailed parameters are listed in Table~\ref{tab:ecf}.

\begin{table*}[htb]
\begin{center}
\begin{threeparttable}
\caption{Energy conversion factors (ECFs) and cross-instrument normalization constants (Referenced to FPM-A)}
\label{tab:ecf}
\begin{tabular*}{\textwidth}{@{\extracolsep{\fill}} lccc @{}}
\hline
Telescope & 
\begin{tabular}[c]{@{}c@{}}ECF\\ (erg cm$^{-2}$ s$^{-1}$ count$^{-1}$)\end{tabular} & 
\begin{tabular}[c]{@{}c@{}}Scaling Constant\\ \end{tabular} & 
\begin{tabular}[c]{@{}c@{}}Corrected ECF\\ (erg cm$^{-2}$ s$^{-1}$ count$^{-1}$)\end{tabular} \\
\hline
NuSTAR (3–20 keV) & 
\begin{tabular}[c]{@{}l@{}}
$6.58 \times 10^{-11}$ (FPM-A) \\
$7.49 \times 10^{-11}$ (FPM-B)
\end{tabular} &
\begin{tabular}[c]{@{}l@{}}
1.00 (FPM-A) \\
$0.99_{-0.30}^{+0.43}$ (FPM-B)
\end{tabular} &
\begin{tabular}[c]{@{}l@{}}
%$\mathbf{6.497 \times 10^{-11}}$ (FPM-A) \\
%$\mathbf{7.488 \times 10^{-11}}$ (FPM-B)
$6.58 \times 10^{-11}$ (FPM-A) \\
$7.59 \times 10^{-11}$ (FPM-B)
\end{tabular} \\
\hline
EP (0.5–3 keV) &
\begin{tabular}[c]{@{}l@{}}
$9.91 \times 10^{-12}$ (FXT-A) \\
$7.76 \times 10^{-12}$ (FXT-B)
\end{tabular} &
\begin{tabular}[c]{@{}l@{}}
$0.75_{-0.16}^{+0.22}$ (FXT-A) \\
$0.74_{-0.16}^{+0.22}$ (FXT-B)
\end{tabular} &
\begin{tabular}[c]{@{}l@{}}
%$\mathbf{1.321 \times 10^{-11}}$ (FXT-A) \\
%$\mathbf{1.051 \times 10^{-11}}$ (FXT-B)
$1.32 \times 10^{-11}$ (FXT-A) \\
$1.05 \times 10^{-11}$ (FXT-B)
\end{tabular} \\
\hline
Swift (0.3–3 keV) &
\begin{tabular}[c]{@{}l@{}}
$3.22 \times 10^{-11}$ (Seq 1) \\
$3.19 \times 10^{-11}$ (Seq 2)
\end{tabular} &
\begin{tabular}[c]{@{}l@{}}
0.74  \\
0.74 
\end{tabular} &
\begin{tabular}[c]{@{}l@{}}
%$\mathbf{4.328 \times 10^{-11}}$ (evt 1) \\
%$\mathbf{4.325 \times 10^{-11}}$ (evt 2)
$4.33 \times 10^{-11}$ (Seq 1) \\
$4.28 \times 10^{-11}$ (Seq 2)
\end{tabular} \\
\hline
\end{tabular*}

\begin{tablenotes}[flushleft]
\item \textbf{Notes.} ECFs convert count rates to observed flux in each instrument's respective bandpass. ECFs were derived from time-integrated spectra using XSPEC fits; scaling constants represent the multiplicative normalization required to match flux levels across instruments, referenced to NuSTAR FPM-A. Corrected ECFs incorporate these scaling factors to allow uniform flux comparison across datasets. Corrected values are used in light curve analysis.
\end{tablenotes}
\end{threeparttable}
\end{center}
\end{table*}

\iffalse
Since EP–FXT flux measurements were conducted in the 0.5–3 keV band rather than our defined 0.3–3 keV band, a bandpass correction was necessary. We calculated the fractional flux contributions ($f$) of the 0.3–0.5 keV interval to the full 0.3–3 keV interval for each Swift–XRT sequence as:
\begin{equation}
f_1=\frac{1.7292\times10^{-12}}{1.3014\times10^{-11}}\approx0.133,
\end{equation}
\begin{equation}
f_2=\frac{2.0438\times10^{-12}}{1.2312\times10^{-11}}\approx0.166.
\end{equation}

Weighted by exposure times ($t_1$, $t_2$), the average fractional flux is:
\begin{equation}
f_w=\frac{f_1t_1+f_2t_2}{t_1+t_2}\approx0.147.
\end{equation}

Thus, the final correction coefficient for converting the EP–FXT flux from the 0.5–3 keV band to the full 0.3–3 keV band is:
\begin{equation}
\text{Correction coefficient}=\frac{1}{1-f_w}\approx1.173.
\end{equation}

By applying these detailed calculations, we have carefully accounted for instrumental scaling differences and corrected the fluxes for bandpass discrepancies, ensuring accurate and meaningful comparisons across detectors.
\fi

\subsection{Scaling Relations} \label{sec:scaling}

Since the key objective of this analysis is to investigate the contribution of HXR from stellar flares on exoplanetary atmospheric escape, precise determination of fluxes across the EUV, SXR, and HXR spectral bands is essential. The scaling relationship between EUV and SXR luminosities is adopted from \cite{Sanz_Forcada_2011}, which is shown in Table \ref{tab:scaling_laws}. We emphasize that the EUV–SXR scaling relation was derived over the full sequence of F through M dwarfs in the XUV wavelength range. AU Mic, being an active M1 star \citep{plavchan2020nature}, falls directly within the spectral class for which the fit was derived. Furthermore, the original sample spans X‑ray luminosities from $\approx10^{26}$ to $10^{31}\,\mathrm{erg\,s^{-1}}$, bracketing AU Mic’s measured quiescent and flaring $L_X$ (on the order of $10^{29}\,\mathrm{erg\,s^{-1}}$). Hence, our use of this relation involves neither extrapolation beyond the fitted domain nor application to atypical activity levels, and is fully justified for estimating the star’s unobserved EUV output.

\begin{figure}[h!]
    \centering
    \includegraphics[width=0.48\textwidth]{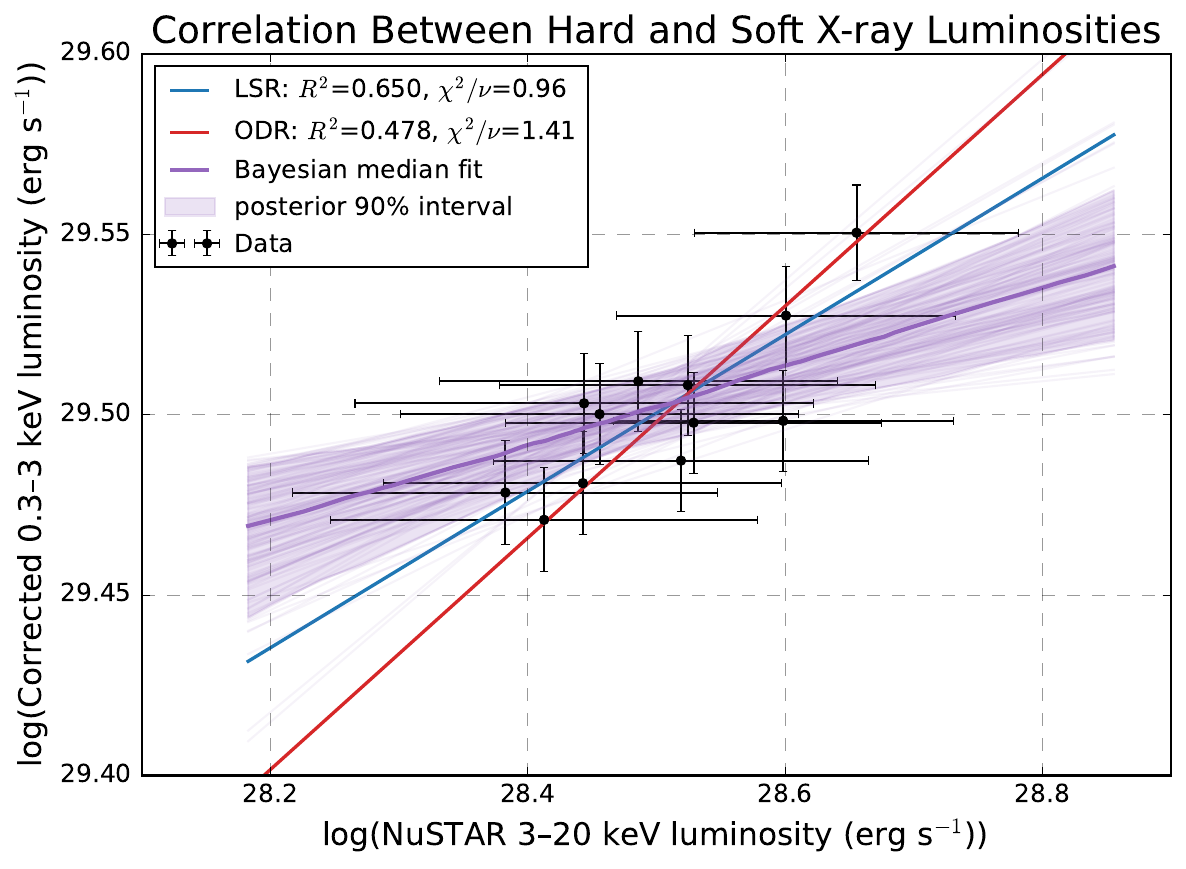} 
    \caption{
    Correlation between HXR (NuSTAR, 3–20 keV) and SXR (corrected EP, 0.3–3 keV) luminosities. 
    The data are binned with a fixed bin size of 240 s to minimize statistical noise from NuSTAR's relatively low count rates.
    NuSTAR time bins with zero count rates were excluded to avoid suppression of correlation strength.
    Additionally, a small number of bins exhibiting anomalously high or low luminosities, often due to edge effects or dominated by upper limits with large uncertainties, were excluded from the regression fits. These excluded points typically had extremely weak or absent hard X-ray signals.
    The blue line shows the best-fit least squares regression (LSR), while the red line represents the orthogonal distance regression (ODR) which accounts for errors in both axes.
    The purple line represents the median of the Bayesian linear regression posterior, obtained via Markov Chain Monte Carlo sampling, with the shaded region showing the 90\% posterior confidence interval.
    Corresponding fit parameters are listed in Table~\ref{tab:scaling_laws}. The SXR–HXR luminosity scaling we derive spans only one order of magnitude in luminosity; expanding the dynamic range is required to generalize the slope.
    }
    \label{fig:hxr_sxr_scaling}
\end{figure}

To establish a complete luminosity profile across all three bands (EUV, SXR, and HXR), we derived a scaling relation between HXR and SXR. Because EP–FXT measurements were originally limited to the 0.5–3 keV band, we applied a bandpass correction to account for the missing 0.3–0.5 keV interval. Based on flux ratios extracted from multiple Swift–XRT sequences and weighted by their respective exposure times, we determined that the 0.3–0.5 keV sub-band contributes $\approx$ 14.7\% of the total 0.3–3 keV flux. This leads to a final correction factor of 1.173.

By applying this correction, we generated accurately corrected EP fluxes for the 0.3–3 keV band, enabling direct comparisons with HXR luminosities. We performed simultaneous measurements of corrected SXR and HXR luminosities over common observational intervals within quiescent periods, deriving scaling relations via both ordinary least squares regression (LSR) and orthogonal distance regression (ODR) \citep{BOGGS1988169}. Given the limited number of available data points and the relatively large measurement uncertainties, we performed an additional generative Bayesian analysis of the scaling relation to assess the robustness of the correlation \citep{hogg2010dataanalysisrecipesfitting}. Using Markov Chain Monte Carlo sampling with the \textit{emcee} package \citep{Foreman_Mackey_2013}, we derived the posterior probability distribution for the slope. The analysis indicates that there is a 99.7\% probability that the slope is positive, providing strong statistical evidence for a positive correlation between the hard and soft X-ray luminosities. However, the precision of this correlation is limited by the current sample size and signal-to-noise ratio; additional observations would both shrink the error bars and extend the dynamic range, enabling a more stringent constraint on the scaling relation. For consistency across the study, we adopt ODR as our primary fitting method, as it accounts for uncertainties in both dependent and independent variables—a critical consideration here given that both SXR and HXR luminosities carry non-negligible errors. The Bayesian posterior median slope and intercept are reported for reference. In contrast, LSR minimizes residuals only in the dependent variable (SXR) and therefore underestimates the slope when HXR uncertainties are significant. The LSR fit is included solely as a consistency check; its agreement with the ODR results (Table~\ref{tab:hxr_ratios}) further supports the robustness of the derived scaling relation. Detailed parameters and regression results are provided in Table \ref{tab:scaling_laws} and Figure \ref{fig:hxr_sxr_scaling}.

\begin{table*}[htb]

\centering

\caption{Summary of scaling relations. All relations are in the form $\log_{10}(L_{\text{Band 1}}) = a \cdot \log_{10}(L_{\text{Band 2}}) + b$}
\label{tab:scaling_laws}
\begin{tabular*}{\textwidth}{@{\extracolsep{\fill}} lccc @{}}
\hline
Relation & Method & Slope $a$ & Intercept $b$ \\
\hline
EUV vs. SXR & Literature \citep{Sanz_Forcada_2011} & $0.860\pm0.073$ & $4.80\pm1.99$ \\
\hline
SXR vs. HXR &  Least Squares Regression & $0.217\pm0.049$ & $23.32\pm1.40$ \\
\hline
SXR vs. HXR & Orthogonal Distance Regression & $0.321\pm0.069$ & $20.35\pm1.97$\\
\hline
SXR vs. HXR & Bayesian Posterior & $0.108^{+0.040}_{-0.033}$ & $26.43^{+0.93}_{-1.15}$\\
\hline
\end{tabular*}
\end{table*}

\subsection{Flares Analysis} \label{sec:flare_analysis}

\subsubsection{Single Temperature Model} \label{sec:spectrum}

\begin{table*}[htb]
  \renewcommand{\arraystretch}{1}
  \small
  \centering
  \begin{threeparttable}
    \caption{Flare parameters from the spectral analysis}
    \label{tab:flare_model}

    \begin{tabular*}{\textwidth}{@{\extracolsep{\fill}} lccc @{}}
      \toprule
      \textbf{Parameters} & \textbf{Unit} & \textbf{TBabs + APEC} & \textbf{TBabs + APEC + power-law}\\
      \midrule
      \multicolumn{4}{c}{\textbf{Flare 1}} \\
      \hline
      $N_\mathrm{H}$\tnote{a}  & $10^{20}$\,cm$^{-2}$
        & $3.9$\tnote{i} & $3.9$\tnote{i} \\
      $T$\tnote{b} 
        & MK & $25.9^{+4.4}_{-3.5}$ & $20.3^{+6.4}_{-4.9}$ \\
      $Z$\tnote{c} 
        &— & $0.15^{+0.19}_{-0.12}$ & $0.26^{+0.68}_{-0.19}$ \\
      ${EM}$\tnote{d}  & $10^{51}\,\mathrm{cm^{-3}}$
        & $9.5^{+3.1}_{-2.3}$ & $12.3^{+6.6}_{-3.9}$ \\ 
      $\Gamma$\tnote{e} &— &— & 2 \tnote{j} \\
      Norm\tnote{f} 
        &$10^{-4}$ & — & $1.3_{-0.6}^{+0.6}$   \\
      $N_{\rm dof}$\tnote{g}
       &— & 200 & 199  \\
      $C_{\rm stat}$\tnote{h}
       &— & 199.4 & 196.3  \\
       Average flux     & $10^{-12}\,\mathrm{erg\,cm^{-2}\,s^{-1}}$      & $1.9_{-0.2}^{+0.2}$       & $2.0_{-0.2}^{+0.2}$    \\
       Average power-law flux     & $10^{-12}\,\mathrm{erg\,cm^{-2}\,s^{-1}}$      & —       & $0.4_{-0.2}^{+0.2}$    \\
    Corrected ECF     & $\mathrm{erg\,cm^{-2}\,s^{-1}\,count^{-1}}$  & \begin{tabular}[t]{@{}l@{}}
                    $6.16 \times 10^{-11}$ (FPM-A)\\
                    $7.09 \times 10^{-11}$ (FPM-B) 
                \end{tabular} 
    & \begin{tabular}[t]{@{}l@{}}
                    $6.45 \times 10^{-11}$ (FPM-A)\\
                    $7.47 \times 10^{-11}$ (FPM-B) 
                \end{tabular} 
                \\
      \midrule
      \multicolumn{4}{c}{\textbf{Flare 2}} \\
      \hline
      $N_\mathrm{H}$ & $10^{20}$\,cm$^{-2}$
        & $3.9$\tnote{i} & $3.9$\tnote{i} \\
      $T$ 
        & MK & $26.2^{+2.9}_{-2.4}$ & $20.8^{+3.7}_{-3.1}$  \\
      $Z$ 
        &—& $0.13^{+0.11}_{-0.09}$ & $0.21^{+0.22}_{-0.13}$  \\
      ${EM}$  
        &$10^{51}\,\mathrm{cm^{-3}}$& $8.4^{+1.8}_{-1.4}$ & $10.6^{+3.5}_{-2.4}$   \\
      $\Gamma$ &— &— & 2\tnote{j}\\
      Norm 
        & $10^{-4}$ & —  & $1.3^{+0.7}_{-0.7}$ \\
    $N_{\rm dof}$
        &—& 300 &299  \\
    $C_{\rm stat}$
        &—& 345.7 &336.1  \\
    Average flux     & $10^{-12}\,\mathrm{erg\,cm^{-2}\,s^{-1}}$      & $1.7_{-0.1}^{+0.1}$       & $1.8_{-0.1}^{+0.1}$    \\
    Average power-law flux     & $10^{-12}\,\mathrm{erg\,cm^{-2}\,s^{-1}}$      & —       & $0.4_{-0.1}^{+0.1}$    \\
    Corrected ECF     & $\mathrm{erg\,cm^{-2}\,s^{-1}\,count^{-1}}$      & \begin{tabular}[t]{@{}l@{}}
                    $6.08 \times 10^{-11}$ (FPM-A)\\
                    $7.00 \times 10^{-11}$ (FPM-B) 
                \end{tabular} 
    & \begin{tabular}[t]{@{}l@{}}
                    $6.40 \times 10^{-11}$ (FPM-A)\\
                    $7.37 \times 10^{-11}$ (FPM-B) 
                \end{tabular} 
       
    \\
  
      \bottomrule
    \end{tabular*}

    \vspace{0.5em}
    \begin{tablenotes}[flushleft]
      \footnotesize
      \item[a] $N_{\rm H}$: hydrogen column density.
      \item[b] $T$: plasma temperature.
      \item[c] Abundance relative to solar.
      \item[d] EM: emission measure, adopting a distance of 9.72 pc \citep{plavchan2020nature}.
      \item[e] $\Gamma$: power‐law photon index.
      \item[f] Power‐law normalization at 1 keV in units of photons keV$^{-1}$ cm$^{-2}$ s$^{-1}$.
      \item[g] $N_{\rm dof}$: the number of independent spectral bins minus the number of free model parameters.
      \item[h] $C_{\rm stat}$: The Poisson‐based maximum‐likelihood statistic used to evaluate fit quality when count rates are low.
      \item[i] The hydrogen column density was fixed to $3.9\times10^{20}\,$cm$^{-2}$, consistent with the value used in Table \ref{tab:specfit}.
      \item[j] The photon index was fixed to 2, consistent with the value used in Table \ref{tab:specfit}.
    \end{tablenotes}

  \end{threeparttable}
\end{table*}

We began the spectral analysis by carefully examining the background spectrum for the NuSTAR observation, ensuring data quality and consistency. To minimize contamination from instrumental background, we ignored energies above 20 keV for each flare, as counts in this range exceeded those from the source background. 

For spectral modeling, we employed the absorbed thermal plasma model \texttt{tbabs*apec}, which is widely used to characterize coronal X-ray emission from stellar flares. Although we also tested a two-temperature thermal model, it did not yield a significant improvement in the fit statistics or reveal a distinct hot component; we therefore adopted a single-temperature model throughout. As NuSTAR is insensitive to soft X-ray absorption due to its lower energy limit of 3 keV, the hydrogen column density $N_\mathrm{H}$ could not be constrained from NuSTAR data alone. Accordingly, we fixed $N_\mathrm{H}$ to the average value derived from contemporaneous Swift-XRT fits.

To assess the potential presence of high-energy emission, we introduced an additional \texttt{power-law} or \texttt{bremss} component \citep{Osten_2007}. For Flare 1, including a \texttt{power-law} component reduced the $\Delta C_{\rm stat}$ by only 3.1, which is not statistically significant given the associated increase in degrees of freedom. The additional \texttt{bremss} component, meanwhile, remained unconstrained. As a result, we concluded that \texttt{power-law} and \texttt{bremss} are not warranted for Flare 1 and adopted the baseline \texttt{tbabs*apec} model. In contrast, for Flare 2, the \texttt{power-law} component yielded a substantial improvement ($\Delta C_{\rm stat} = 9.6$), exceeding the $\approx$4.6 threshold for 90\% significance for two additional parameters, and thus suggesting the presence of a possible nonthermal tail. Due to limited signal-to-noise at high energies, the photon index could not be constrained and was therefore fixed to 2, a typical value for stellar flare spectra \citep{Benz2017}.  Additionally, when the temperature of the \texttt{bremss} component is left free, it reaches the upper bound (200 keV), suggesting a poor constraint. However, fixing the temperature to 25 keV yields a comparable statistical improvement to the \texttt{power-law} fit ($\Delta C_{\rm stat} = 9.5$), indicating that a high temperature thermal component remains a viable explanation for the observed hard excess. In Section \ref{sec:HEtail}, we discuss the plausibility of both models, and ultimately adopt the \texttt{tbabs*(apec+power-law)} model for Flare 2.

The best-fit parameters for each flare and spectral model are summarized in Table~\ref{tab:flare_model}, while the detailed parameters for the \texttt{bremss} component in Flare 2 are listed separately in Table~\ref{tab:bremss}.

\subsubsection{6.4 keV Emission Line}
\begin{figure}[ht!]
  \centering
  \includegraphics[width=\columnwidth]{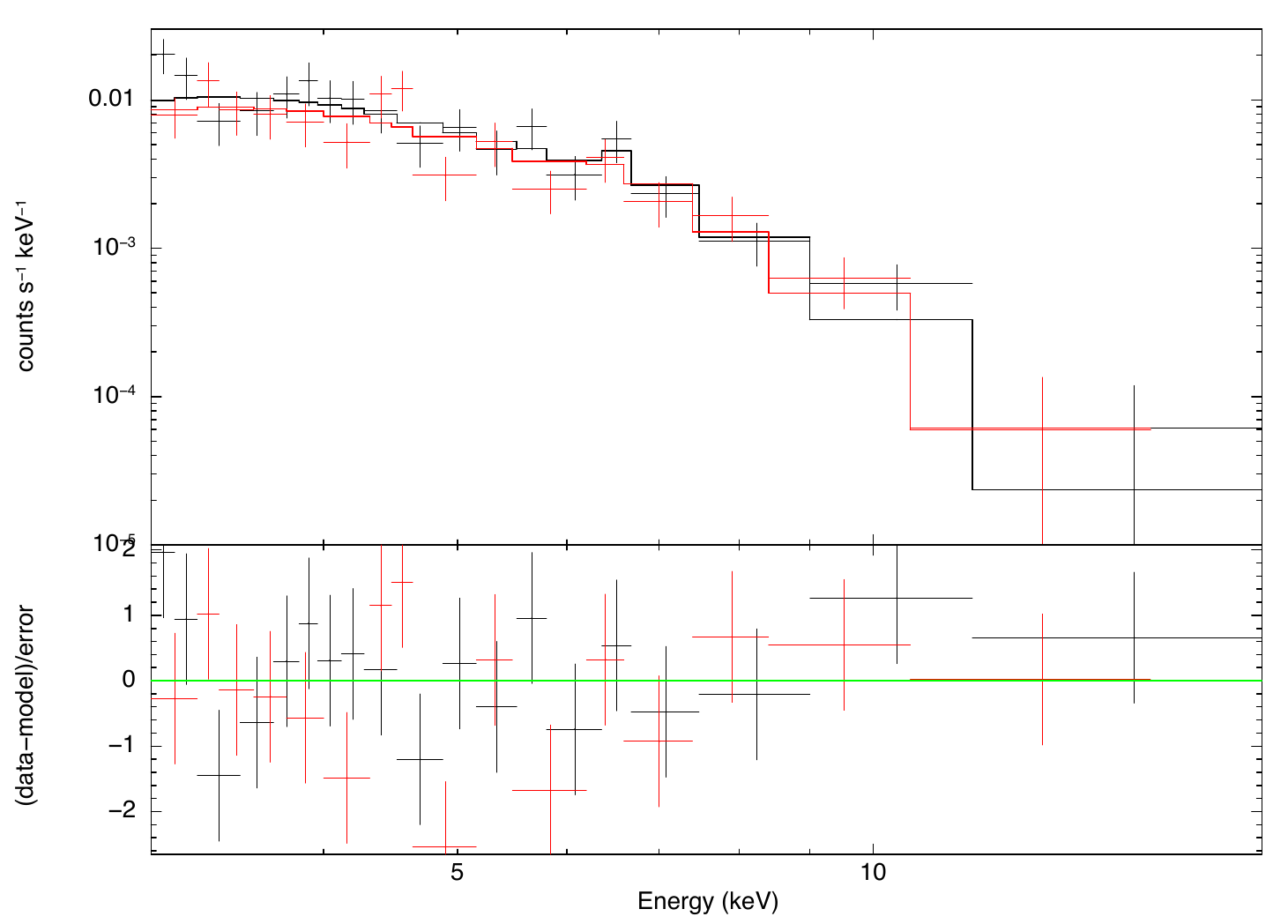}
  
  \vspace{1em}
  % Flare 2 spectrum
  \includegraphics[width=\columnwidth]{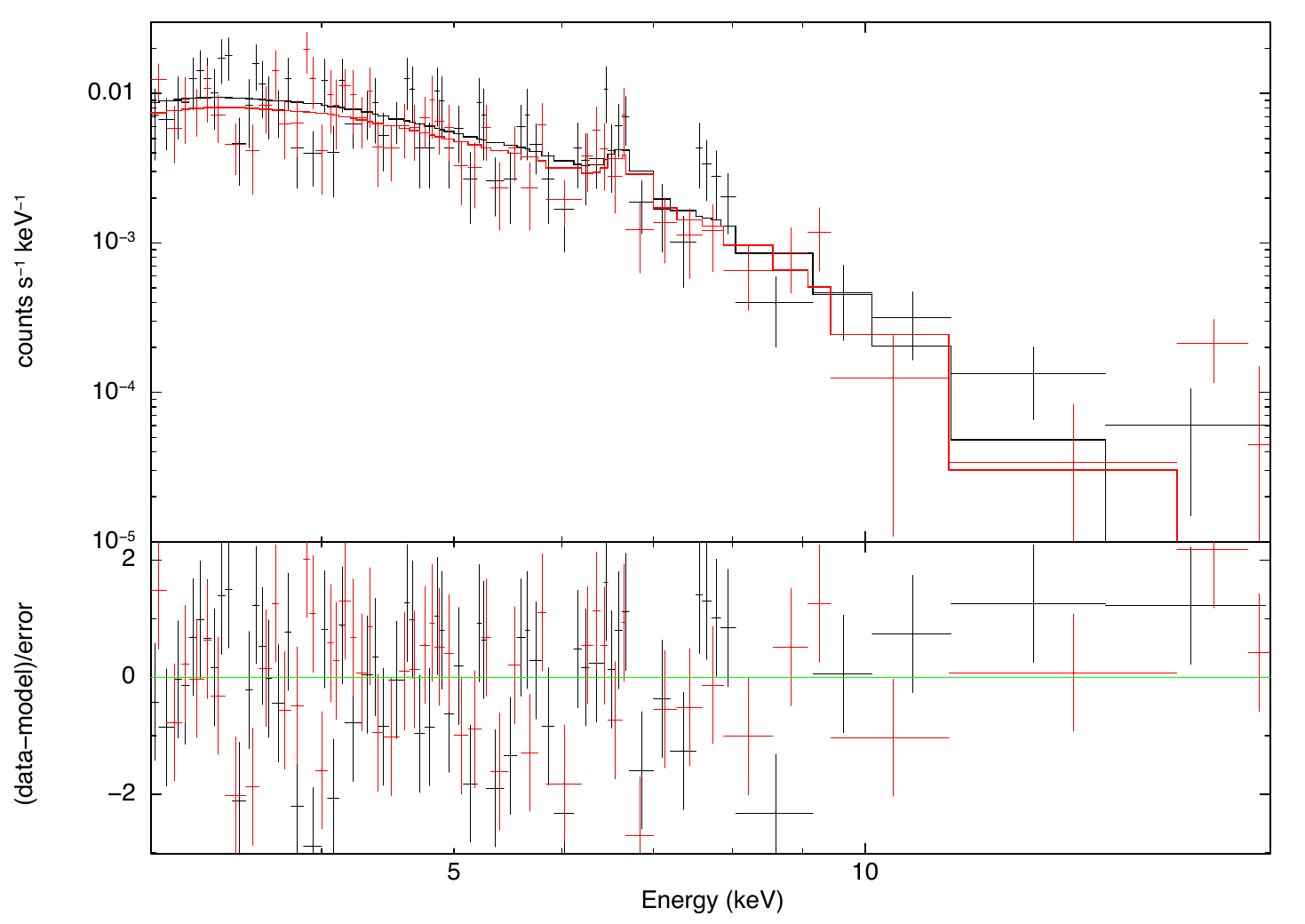}
  
  \caption{Top: NuSTAR spectrum of Flare 1 with the best‐fit absorbed \texttt{apec} model overlaid. Bottom: NuSTAR spectrum of Flare 2 with the best‐fit absorbed \texttt{apec} model overlaid. Black points denote FPM-A data and red points denote FPM-B data.}
  \label{fig:flare_spectra}
\end{figure}

After fitting each flare with the absorbed \texttt{apec} model, which already reproduces the thermal Fe XXV complex near 6.7 keV, we found that the second flare exhibited additional flux in the 6.3–6.5 keV range (Figure \ref{fig:flare_spectra}). To model this excess, we added a narrow Gaussian component fixed at \(E = 6.40\)\,keV with \(\sigma = 0.10\)\,keV, leaving only its normalization free. Because our energy bins span both the 6.4 keV and 6.7 keV features and NuSTAR’s resolution cannot fully separate them, the added line’s flux inherently trades off against the \texttt{apec} abundance in the fit.

The resulting equivalent width (EW) shows a marginal but not significant improvement in fit statistic when the Gaussian is included for Flare 2. There is also no significant change for Flare 1. Consequently, while the addition of the 6.4 keV component hints at fluorescence during the second event, it is not statistically significant in either flare. 

\subsubsection{Energy Ratio} \label{sec:energy_ratio}
\begin{figure}[h]
  \centering
  \includegraphics[width=\columnwidth]{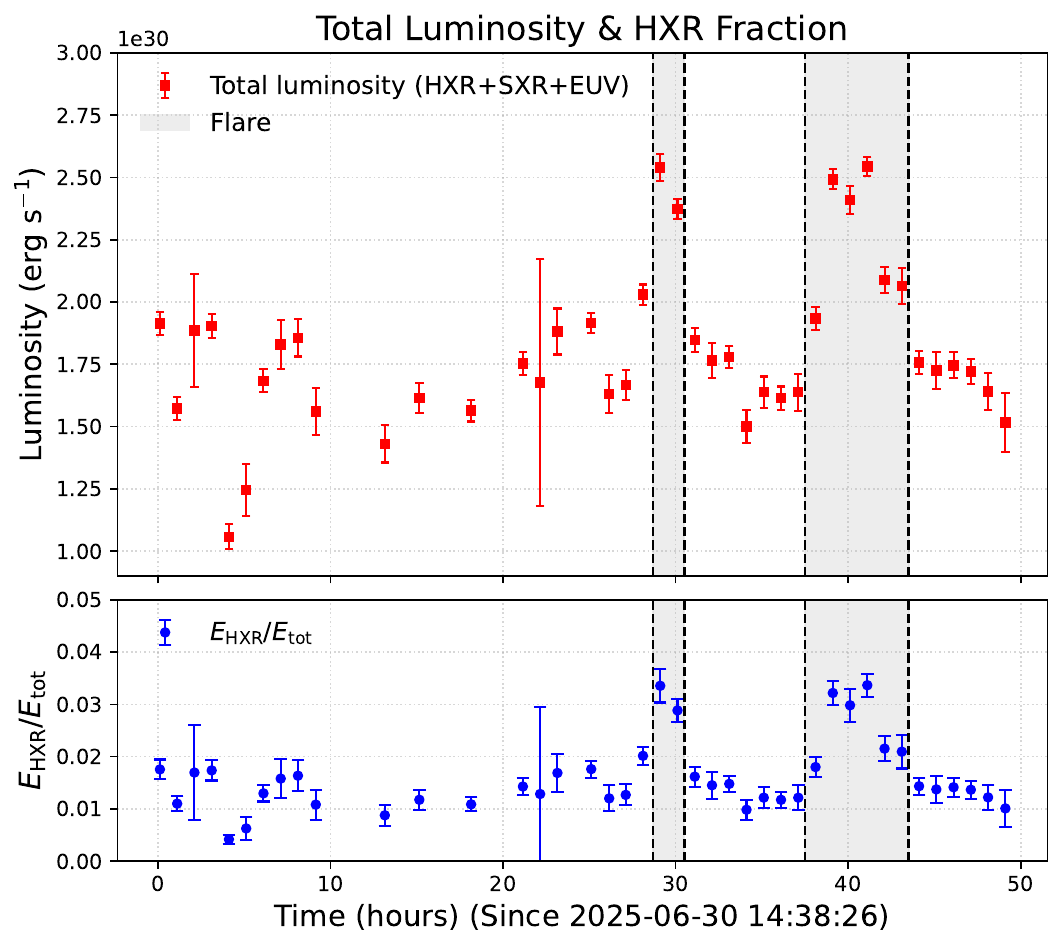}
  \caption{Full band total luminosity and HXR energy fraction (bin size=3600 s). Top: Total luminosity $L_{\mathrm{tot}} = L_{\mathrm{HXR}} + L_{\mathrm{SXR}} + L_{\mathrm{EUV}}$ obtained using ODR scaling relations. Error bars show $1\sigma$ uncertainties. Shaded grey bands mark flare intervals. 
  Bottom: Time evolution of the HXR energy fraction $E_{\mathrm{HXR}} / E_{\mathrm{tot}}$, where energies are per-bin integrals $E = L\,\Delta t$ with $\Delta t = 3600~\mathrm{s}$. Results use the same ODR relations; error bars denote $1\sigma$ uncertainties in $E_{\mathrm{HXR}} / E_{\mathrm{tot}}$. The fraction stays at the few-percent level in quiescence and modestly increases during flares, indicating only limited changes in the spectral energy partitioning across HXR, SXR, and EUV.}
  \label{fig:hxr_fraction_time}
\end{figure}

To quantify the contribution of HXR emission to exoplanetary atmospheric escape, it is essential to accurately determine the relative energy distribution among the HXR, SXR, and EUV spectral bands. Utilizing the scaling relations derived in Section \ref{sec:scaling}, we note that these SXR–HXR and SXR-EUV conversions were originally calibrated on quiescent coronal observations and hence carry additional systematic uncertainty when extrapolated to impulsive flare, state plasmas, and remain consistent with thermal-dominance ratios from solar and stellar studies, a robustness we further discuss in Section \ref{sec:uncertainties}. The ECFs in Table \ref{tab:ecf} and \ref{tab:flare_model}, and the flare spectral fitting in Section \ref{sec:spectrum}, we computed the luminosities in the SXR and HXR bands, subsequently estimating EUV luminosities via the established scaling relationship from \cite{Sanz_Forcada_2011}. The relative energy contributions of HXR within the combined HXR+SXR+EUV bands were thus determined for the two distinct flare intervals as well as for the global average, facilitating a clear assessment of the energetic significance of HXR during flare events. The comparative results using LSR, ODR, and Bayesian posterior methods are summarized in Table~\ref{tab:hxr_ratios}, showing that the inferred HXR fractions at flare peaks are robust across all fitting approaches. Although the lower slopes from the LSR fit and the Bayesian posterior tend to yield slightly higher HXR fluxes, the resulting peak fractions differ by no more than 1\%. The fraction of HXR energy to the total radiative energy as a function of time is shown in Figure \ref{fig:hxr_fraction_time}.

\begin{figure}[h]
    \centering
    \includegraphics[width=\columnwidth]{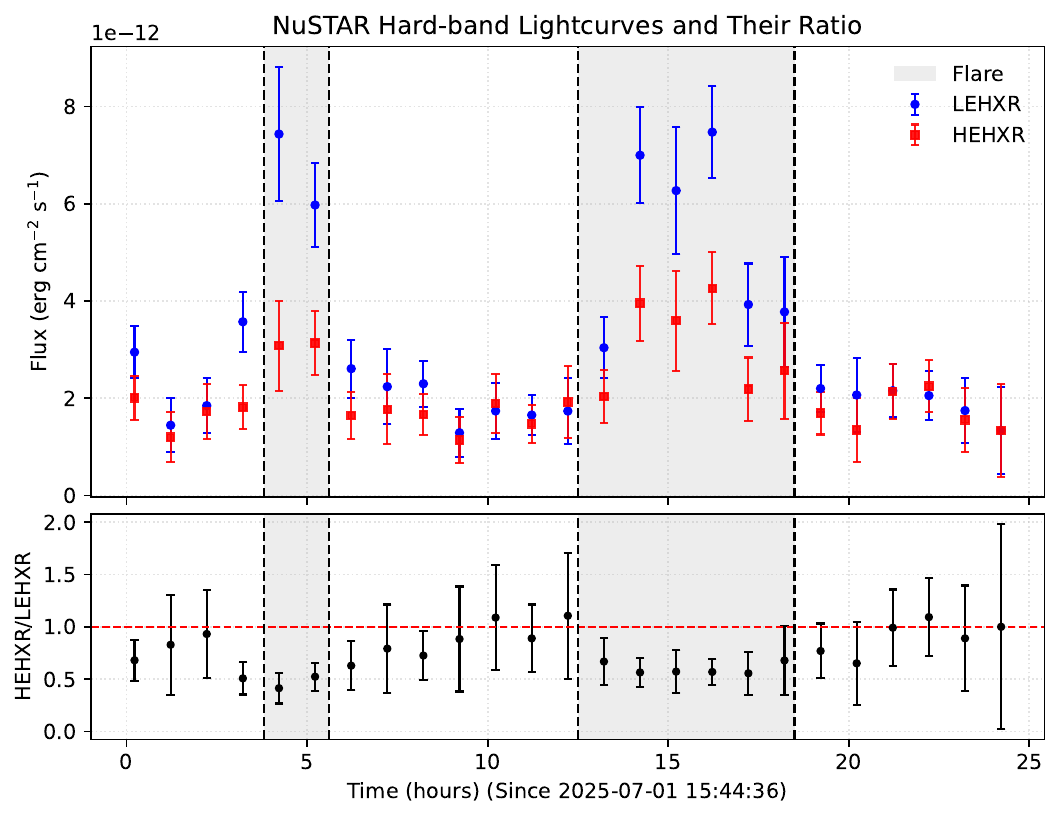}
    \caption{NuSTAR hard X-ray light curves and hardness evolution. Top: energy-flux light curves in 3–6 keV (LEHXR, blue circles) and 6–20 keV (HEHXR, red squares) obtained by converting FPMA/B count rates using spectral-fit ECFs and averaging the two modules; points show 1$\sigma$ uncertainties. Grey shaded bands mark the identified flare intervals. Bottom: time-resolved fraction of HEHXR relative to the LEHXR; the red dashed line denotes equal contributions from low and high bands. The HEHXR fraction decreases during flares, indicating that the flare-associated enhancement is concentrated in the lower-energy 3–6 keV range. Bin size is 3600 s.}
    \label{fig:hxr_fraction}
\end{figure}

Furthermore, we subdivided the HXR band into two energy intervals: low-energy HXR (LEHXR, 3–6 keV) and high-energy HXR (HEHXR, 6–20 keV), and performed spectral fitting for each sub-band, converting the results into fluxes. Analysis revealed that during flare processes, the fractional contribution of the HEHXR component to the total HXR luminosity decreases, as illustrated in Figure \ref{fig:hxr_fraction}. This indicates that the enhancement in total HXR emission during flares predominantly arises from LEHXR photons, which is consistent with softening due to dominance of thermalized plasma rather than sustained high-energy acceleration, providing deeper insight into the flare-induced energy redistribution in stellar coronae.

\begin{table*}[t]
\centering
\caption{HXR fraction of the total flare energy\tnote{a} estimated using different methods}
\label{tab:hxr_ratios}
\begin{threeparttable}
\begin{tabular*}{\textwidth}{@{\extracolsep{\fill}} lccc @{}}
\toprule
\textbf{Event} & \textbf{LSR} & \textbf{ODR} & \textbf{Bayesian Posterior} \\
\midrule
Average   & $0.018^{+0.003}_{-0.002}$ & $0.017^{+0.002}_{-0.003}$ & $0.016^{+0.002}_{-0.002}$\\
Flare 1   & $0.033^{+0.004}_{-0.005}$ & $0.033^{+0.005}_{-0.005}$ & $0.041^{+0.006}_{-0.005}$\\
Flare 2   & $0.034^{+0.006}_{-0.006}$ & $0.033^{+0.004}_{-0.004}$ & $0.041^{+0.004}_{-0.004}$ \\
\bottomrule
\end{tabular*}
\begin{tablenotes}
\item[] \textbf{Notes.} Total refers to the sum of the energies in the EUV, SXR, and HXR bands.
\end{tablenotes}
\end{threeparttable}
\end{table*}

\subsubsection{GOES Class}

The AU Mic flares rival the most extreme solar events.  Solar‐flare magnitudes are traditionally expressed in terms of the 1–8 \AA~SXR peak flux measured by the Geostationary Operational Environmental Satellite (GOES), with classes A, B, C, M, and X corresponding to peak flux thresholds of \(10^{-8}\), \(10^{-7}\), \(10^{-6}\), \(10^{-5}\), and \(10^{-4}\)\,W\,m\(^{-2}\) respectively.  From our isothermal \texttt{apec} fits, characterized by temperature \(T\) and emission measure \(\mathrm{EM}\), we synthesize the 1–8 \AA~flux at a distance of 1 AU using\cite{PIMMS2025}. Converting our flare parameters to an equivalent GOES peak flux yields classes of approximately X700–X1400. By comparison, the largest recorded solar flare \citep{kane2005multispacecraft} corresponds to an X28 event in November 2003.  Hence, the AU Mic flares studied here are \(\approx\!25\!-\!50\times\) more energetic than any solar flare on record.

\subsubsection{Chromospheric Evaporation Model}\label{sec:cem}

\begin{figure}[h!]
\centering
\includegraphics[width=\columnwidth]{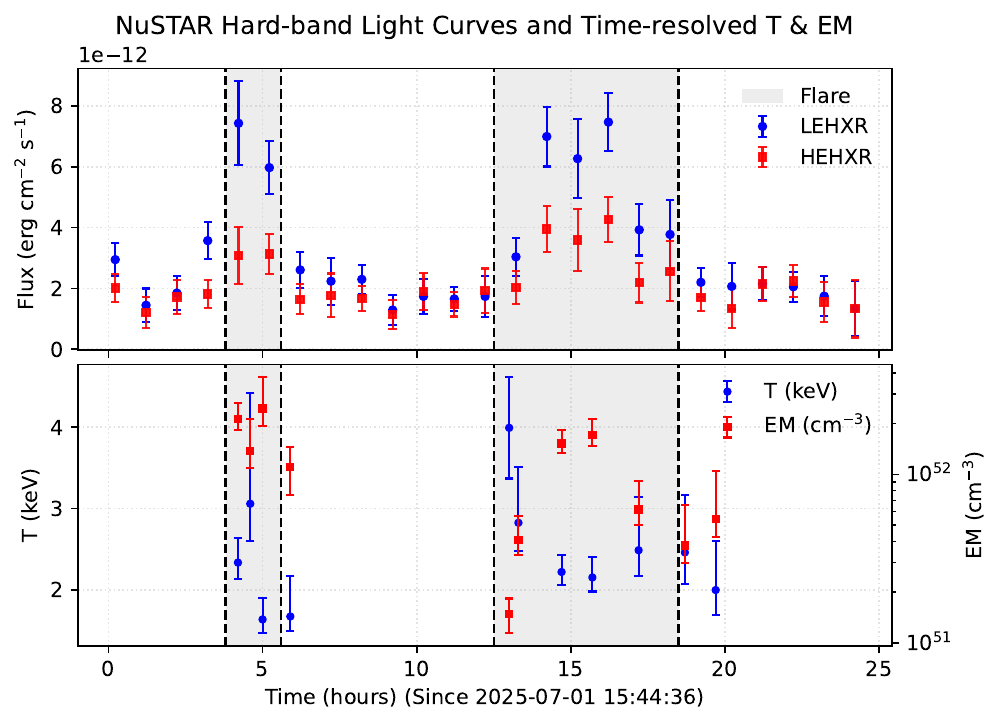}
\caption{
NuSTAR hard-band light curves and time-resolved plasma diagnostics (bottom). The lower panel shows the best-fit temperature T (left axis) and emission measure EM (right axis) with asymmetric error bars. Note that the bin sizes for T and EM differ: for the first flare we use 1200~s bins; for the second flare we use per-orbit bins, except the early rise which uses 1800~s. Shaded regions indicate flare intervals; time is in hours since the reference epoch on the x-axis.
}
\label{fig:t_em}
\end{figure}

To trace the thermodynamic evolution of the two major flares, we performed time-resolved spectroscopy with phase segmentation guided by photon statistics. For Flare 1 we extracted a rise, two exponential decay segments, and quiescence; for Flare 2 we split the rise into two sub-intervals and then used per-orbit bins through the decay. Bin sizes were 1200 s for Flare 1 and 1800 s for the early rise of Flare 2; finer slicing was avoided due to limited counts.

We jointly fit FPMA/FPMB spectra with \texttt{tbabs*apec}, fixing $N_{\rm H}$ to the Swift value to reduce degeneracy. W-statistics with minimal grouping were applied, and asymmetric uncertainties reported. Adding a power-law component did not materially affect the derived $T$ and $EM$.

Both flares showed consistent behavior: $T(t)$ rose rapidly and peaked before $EM(t)$, which then lagged and declined alongside $T$ during exponential cooling. Hard-band light curves exhibited hard-to-soft evolution through the peak and early decay. The lag between $T$ and $EM$ maxima was constrained only to a fraction of the bin size but was robust to re-binning.

The $T$–$EM$ trajectory matches standard chromospheric evaporation: impulsive heating drives a rapid $T$ rise; evaporation increases $EM$ with a delay; both then decrease as the loop cools by conduction and radiation \citep{1974SoPh...34..323H, Shibata_2002}. While this $T$-first, $EM$-delayed sequence alone is not a unique proof, the observed ordering and spectral hard-to-soft trend support a thermally dominated flare evolution.

\section{Discussion} \label{sec:discussion}
Spectral analysis of two NuSTAR‐observed flares, each spanning roughly 3--10 ks, reveals hot, thermal plasma with peak temperatures of 20--30 MK. An absorbed \texttt{apec} model alone provides an excellent fit in both cases. For Flare 1, adding a \texttt{power-law} component yielded only a marginal improvement, while a \texttt{bremss} component remained unconstrained. This indicates a lack of statistically significant evidence for either a broadband nonthermal tail or high temperature thermal emission. However, Flare 2 showed a more significant improvement in fit statistics, and residuals above 10 keV suggest the presence of a high-energy tail. 
%This flare also exhibits a localized Neupert effect in each of the two rise phases, indicative of impulsive particle acceleration and further supporting the presence of a nonthermal component. 
The \texttt{bremss} model yields a comparable statistical improvement to the \texttt{power-law} model, indicating that a high temperature thermal emission is also a viable explanation. Notably, although the second event exhibits a nonzero equivalent width for a narrow 6.4 keV Fe K$\alpha$ emission line, there is no statistically significant evidence for iron fluorescence. This is because the uncertainties on the equivalent width are estimated at the 90\% confidence level, whereas a $3\sigma$ threshold (99.7\%) is typically required to claim a detection; under such a threshold, the equivalent width becomes unconstrained.

For comparison, our measured temperatures lie at the upper end of the distribution reported in previous surveys. \cite{Tristan_2023} analyzed a large flare sample on AU Mic with XMM--Newton and found peak temperatures of 10--30 MK. This agreement is likely due to the hard X-ray sensitivity of NuSTAR, which allows measurements of hot flare plasma components that may be underconstrained in softer bands. Similarly, \cite{Mitra_Kraev_2005} fitted a three–temperature collisional ionization equilibrium (CIE) model to EPIC‐PN spectra over 0.2--12 keV and reported a total 0.2--12 keV flare luminosity of $2.99\times10^{29}$\,erg\,s$^{-1}$, consistent with our integrated luminosities adopting a comparable bandwidth. 

\subsection{High-Energy Tail} \label{sec:HEtail}

Our spectral analysis for Flare 2 reveals a potential high-energy tail above 10 keV, which is better fit with an additional component beyond a simple thermal model. Two possible interpretations were explored: a nonthermal power-law component representing accelerated electrons, and high temperature thermal bremsstrahlung emission.

From a physical perspective, maintaining such a hot plasma (25 keV $\approx$ 290 MK) would demand an extraordinarily high energy budget. For large stellar flares, the total thermal energy required can be estimated as:
%\begin{equation}
%E = 3 k_{\mathrm{B}} T \times \sqrt{\mathrm{EM} \times V},
%\end{equation}
\begin{equation}
{E = 3 k_{\mathrm{B}} T \, \frac{{EM}}{n_{\mathrm{e}}},}
\end{equation}
where $k_{\mathrm{B}}$ is Boltzmann’s constant and $n_{\mathrm{e}}$ is the electron number density. The electron number density of AU Mic is 2--5$\times$$10^{12}\,$cm$^{-3}$ from Fe\,\textsc{xxi-xxii} \citep{Schrijver1995AandA302,Gudel:2004bz}, leading to an energy estimate on the order of $10^{32-33}$ erg. As shown by solar-flare energetics scalings \citep{Cliver_2022,Shibata2013PASJ65}, flares of GOES class $\approx$X1000 typically reach total radiated energies of $\approx10^{34}$~erg. The events analyzed here are of comparable magnitude, so an energy scale of $\approx10^{34}$~erg is consistent with these relations and does not imply a contradiction with the thermal energy we calculated. Importantly, the presence of an extremely hot thermal plasma near the peak remains plausible and cannot be ruled out by our data.

When modeling the 3--20~keV spectra with a thermal component plus a power-law, the power-law contribution to the model flux remains roughly constant at $\approx$20\% across both quiescent and flare intervals. This percentage is derived from time-resolved fits in which, for each time bin (including quiescent and flare sub-intervals), we compute the power-law flux divided by the total model flux; across all bins the share clusters near 20--25\% with no significant trend once statistical errors are propagated. We stress, however, that within our limited bandpass a very hot thermal continuum can mimic a power law over a restricted energy range, so this fraction is inherently model-dependent and does not by itself establish a non-thermal origin for the high-energy tail.

A plausible interpretation consistent with the near-constant fraction is that the corona hosts a quasi-steady (statistically steady but temporally intermittent) population of accelerated electrons maintained by numerous small-scale reconnection events or turbulence-driven acceleration. In this picture, major flares chiefly enhance the thermal emission via chromospheric evaporation and loop filling, rather than substantially increasing the efficiency of particle acceleration. The observed reduction in the 6--20~keV fraction during flares can then be understood as spectral softening caused by preferential heating at lower energies, which dilutes the apparent hard tail. While this behavior is compatible with evaporation-driven heating and loop filling (thereby boosting $<\!6$~keV emission), it does not rule out particle acceleration; given our effective sensitivity only up to $\approx$20~keV and background limitations, non-thermal signatures peaking above this range could remain undetected.

An alternative, equally viable explanation for the high-energy tail is an extreme hot thermal component. Over the 3--20~keV band, a plasma with $kT\!\gtrsim\!8$--12~keV (and, in extreme cases, approaching a few $\times 10^8$~K) yields a bremsstrahlung continuum that can appear locally power-law–like, producing fits statistically comparable to thermal+power-law models. In this scenario, the near-constant hard-tail fraction simply reflects a relatively stable hot end of the thermal distribution superposed on a cooler, more variable bulk component. The flare-associated ''softening'' then arises naturally as loop filling increases the emission measure at lower temperatures, making the ultra-hot tail comparatively less prominent. Energetically, with representative parameters, the instantaneous thermal energy remains well below the total flare energy budget inferred from global scalings, and thus an extreme hot component is physically tenable. Decisive spectroscopic anchors (e.g., Fe\,\textsc{xxv}/\textsc{xxvi} around 6.7/6.97~keV) are not detected at sufficient signal-to-noise ratio (SNR) to confirm or refute this interpretation, and the current bandpass does not reach high enough energies to identify a thermal exponential cutoff.

Taken together, our observations favor a picture in which the corona may host a quasi-steady population of accelerated electrons from numerous small-scale reconnection episodes, or an extreme hot thermal tail that is difficult to distinguish from a power-law within 3--20~keV. The flares in our dataset appear predominantly thermally powered—with hard-to-soft evolution and delayed $EM$ rises consistent with chromospheric evaporation, while any localized or weak enhancements in particle acceleration are not unambiguously captured by the current bandpass. Disentangling these scenarios will require higher SNR coverage extending well beyond 20--30~keV to detect a thermal cutoff or a sustained non-thermal extension, firmer Fe\,\textsc{xxv}/\textsc{xxvi} constraints, and coordinated multiwavelength diagnostics to quantify the thermal–non-thermal energy balance in stellar flares.

\subsection{Scaling Relations}\label{sec:uncertainties}

To place the 3--20~keV measurements in a broader radiative context, we estimate SXR and EUV outputs by chaining empirical scaling relations (HXR to SXR and SXR to EUV). We therefore summarize the dominant uncertainties, indicate when the scalings are most reliable, and compare the reconstructed fluxes with independent expectations.

HXR varies more rapidly than SXR and EUV, and unequal binning can bias instantaneous ratios; we mitigate this by using the same time segmentation as the spectral fits and by phase–averaging where needed. Differential Emission Measure (DEM) shape, evolving temperatures, and abundances introduce factor–of–few scatter in EUV reconstructions even when SXR is well constrained. Differences between datasets used to calibrate the relations and ours add further systematics, especially in EUV.

In quiescence, emission is predominantly thermal and multi–temperature, making SXR–EUV mappings closer to their calibration regime. During flares, the plasma follows a chromospheric–evaporation sequence (temperature peaks first, emission measure peaks later), and the DEM evolves rapidly; in this phase, treating all HXR as purely thermal can overstate SXR/EUV if transient non–thermal power is present. We therefore anchor the scaling at each time bin to the phase–resolved thermal parameters ($T$, $EM$) and propagate asymmetric uncertainties.

Where possible, we compare the scaled ratio of SXR and EUV outputs with expectations from solar and stellar flare energetics \citep{kowalski2024stellarflares}. Within uncertainties, the reconstructed SXR and EUV are broadly consistent with these independent estimates, for both quiescent intervals and the main flare phases.

Scaling relations, when used with phase–matched binning and explicit uncertainty propagation, provide a reasonable consistency check on flare energetics across HXR–SXR–EUV bands. We emphasize that the resulting SXR/EUV numbers are not direct observables; they are scaling relation–derived estimates subject to the assumptions outlined above and are interpreted accordingly throughout.

\subsection{Atmospheric Escape Effect}

In this part, we quantify the contribution of HXR emission to exoplanetary atmospheric escape and compare it against the dominant XUV-driven mechanism. Stellar XUV photons predominantly deposit energy in the upper atmosphere via photoionization and subsequent photochemical reactions, producing photoelectrons that efficiently heat the gas and drive hydrodynamic outflows with empirically determined escape efficiencies of $\eta_{\rm XUV}\approx10\text{--}30\%$ \citep{lammer2003, Murray_Clay_2009, Owen_2012}. These photons also play a key role in atmospheric photochemistry by initiating ion–molecule and dissociation reactions that regulate the abundances of species such as H, O, C, and N, and drive catalytic cycles involving HOx and NOx that can significantly alter ozone and other UV-shielding molecules \citep{Segura2010}.

In contrast, HXR photons penetrate to much greater column densities, depositing energy through Compton heating cascades and secondary-electron production deep within the atmosphere \citep{Gudel:2004bz}. Energy deposited at these depths is subject to rapid radiative cooling and conductive losses, making HXR-driven heating intrinsically less effective at powering atmospheric escape. Moreover, because the deposition occurs well below the primary photodissociation and ionization layers, the chemical impact of HXR photons is shifted toward denser, less UV-active regions of the atmosphere, where reaction timescales are slower and photochemical feedbacks are weaker. In such layers, HXR-driven chemistry may enhance the production of certain radicals (e.g., HOx, NOx) through collisional ionization cascades, but these species are less likely to propagate into the upper atmosphere to influence global photochemistry or ozone balance on short timescales.

In the absence of direct measurements of HXR-specific escape efficiencies, and given that the HXR energy deposition mechanism does not couple to the upper atmospheric layers as directly as XUV photoionization, we conservatively adopt the well-characterized XUV-driven efficiency as an upper bound for HXR-induced escape. Denoting the integrated HXR flare energy from Section~4.1 as $E_{\rm HXR}$, the maximal HXR-driven mass-loss rate \citep{CHADNEY2015357} may then be expressed as
\begin{equation}
    \dot{M}_{\rm HXR, max} \propto \frac{\eta_{\rm XUV}{\pi}R_{\rm p}^{2}F_{\rm HXR}}{K\,\bigl(GM_{\rm p}/R_{\rm p}\bigr)},
    \label{eq:hxr_escape}
\end{equation}
where $K\approx1$ is the gravitational potential correction factor and $GM_{\rm p}/R_{\rm p}$ represents the planetary potential well depth.

For a given exoplanetary system, all parameters in Equation
\ref{eq:hxr_escape} remain fixed except for the integrated flare flux $F$ and its associated escape efficiency. Therefore, the maximum HXR-driven mass-loss rate can be compared directly to the XUV-driven rate by forming the ratio
\[
\frac{\dot{M}_{\rm HXR, max}}{\dot{M}_{\rm XUV}} \approx \frac{\eta_{\rm XUV} \;E_{\rm HXR}}{\eta_{\rm XUV} \;E_{\rm XUV}} = \frac{E_{\rm HXR}}{E_{\rm XUV}}.
\]
For the two flares analyzed here, this ratio is $3-4\%$, demonstrating that HXR-induced heating contributes only a small fraction of the total flare-driven atmospheric escape.

In Section \ref{sec:spectrum} and \ref{sec:energy_ratio}, we demonstrate that although magnetic reconnection and associated fast shocks can transiently enhance HXR emission at the flare peak, the integrated energy fraction in the 6--20\,keV band actually declines during the flare, while the 3--6\,keV contribution increases. This behavior indicates that the bulk of flare-induced heating resides at LEHXR rather than in the high-energy nonthermal tail. Consequently, even under the optimistic assumption $\eta_{\rm HXR}=\eta_{\rm XUV}$, HXR photons could contribute at most a few percent of the total flare-driven mass loss. In reality, HXR-driven escape efficiencies are expected to be orders of magnitude lower, justifying the neglect of HXR heating in global atmospheric evolution models. 

From a photochemical perspective, HXR photons may still influence lower atmospheric chemistry through deep ionization, altering the abundances of long-lived species such as CO, CH$_4$, and NH$_3$, and enabling radical-driven catalytic cycles that can indirectly affect stratospheric composition over extended timescales. However, the lack of direct coupling to the primary escape-driving layers and the dominance of rapid cooling processes likely make such effects secondary compared to XUV-driven photochemistry.

While our analysis has focused on the XUV--HXR regime, extensive solar and stellar studies demonstrate that the flare bolometric energy is often dominated by emission from the near-ultraviolet (NUV) through optical bands rather than by X-rays. For the Sun, white-light and NUV continuum account for a large fraction of the radiative output \citep{Kretzschmar_2010,Kretzschmar_2011}, and similar results have been found in stellar flares \citep{France_2013}. In terms of photon numbers, far-ultraviolet (FUV) and NUV emission can exceed HXR photons by many orders of magnitude, exerting a proportionally larger influence on lower-atmosphere heating and chemistry through processes such as photoionization and photodissociation \citep{Segura2010}. Physically, FUV photons are particularly effective at dissociating molecules such as H$_2$O, CO$_2$, and CH$_4$, thereby altering O- and OH-driven chemical pathways that regulate ozone and hydrogen escape, whereas NUV continuum absorption deposits energy deeper than EUV, at pressures of $\approx$0.1--1 bar, i.e., the levels most relevant to climate and surface habitability.

Observations of stellar superflares further support this picture, with NUV fluxes reported to increase by factors of up to $10^{4}$ \citep{kowalski2024risingnearultravioletspectrastellar}, highlighting their potentially dominant role in shaping atmospheric structure and evolution. Because our analysis does not explicitly quantify FUV and NUV deposition, the relative role of HXR in our results is likely overstated when considered in this broader multiwavelength context.

This analysis nevertheless constrains the contribution of hard X-rays to atmospheric escape. For the two flares examined here, even under the upper-limit assumption that the HXR efficiency equals the XUV efficiency, the maximum HXR-driven mass-loss rate, taken to be proportional to the HXR flux, reaches only $\approx$3--4\% of the EUV- and SXR-driven value, underscoring the minor role of HXR heating in atmospheric erosion. Future observational campaigns and laboratory experiments aimed at measuring HXR, NUV, and FUV heating and escape efficiency, such as simultaneous high-energy transit spectroscopy with missions like NewAthena or dedicated laboratory Compton heating experiments on planetary analogs, are essential to constrain this poorly understood yet potentially important regime.

\section{Summary and Conclusions} \label{sec:summary}

Stellar flares represent critical episodes of enhanced high‐energy radiation that can profoundly impact exoplanetary atmospheres through heating and mass loss. In this study, we have analyzed two carefully selected flares with high‐quality multiwavelength coverage, combining scaling laws to extrapolate the integrated energies in the EUV, SXR and HXR bands.

%Time-resolved spectroscopy and flux decomposition reveal that HXR emission from AU Mic contributes only marginally to the overall radiative budget. The HXR fraction increases from $\approx\,$1.7\% in quiescence to $\approx\,$3.3\% at flare peaks, coinciding with nonthermal signatures such as high-energy power-law tails and two distinct localized Neupert effect episodes. However, the 6--20\,keV flux fraction declines during flares, indicating that thermal plasma heating dominates the overall flare energetics.

Time-resolved spectroscopy and flux decomposition show that hard X-rays contribute only marginally to AU~Mic's radiative budget, in both quiescent and flaring conditions, based on scaling relation-derived estimate subject to the stated assumptions. During flares the 6--20~keV flux fraction declines and the $T$--$EM$ sequence follows the chromospheric-evaporation trend, indicating that thermal plasma heating dominates the overall energetics.

By adopting a conservative XUV-based escape efficiency, we estimate that HXR-powered atmospheric escape contributes at most a few percent of the total XUV-driven mass-loss rate. This conclusion holds over the parameter range sampled here (flare temperatures $\approx\,$20--30\,MK, HXR energies $\lesssim$ a few $\times10^{33}$\,erg), and may break down in the case of rare superflares. These results support a scenario in which nonthermal electrons may contribute episodically to chromospheric heating but do not dominate atmospheric escape in AU Mic–like systems; instead, XUV irradiation remains the primary driver of exoplanetary mass loss. A key uncertainty remains the lack of direct measurements for HXR-specific escape efficiencies. To better constrain this poorly understood regime, future work should focus on (1) targeted multiwavelength transit spectroscopy with next-generation X-ray observatories and (2) laboratory or numerical studies of Compton-driven heating in planetary analog atmospheres.

We note that expanding the database to a larger range of flare magnitudes and stellar types will be essential to confirm these conclusions and to explore any variability in high‐energy escape processes. In particular, increasing the number of high‐quality multi-wavelength flare observations will reduce statistical uncertainties and extend the dynamic range, enabling more reliable scaling relations and a clearer determination of the physical connection between SXR and HXR emission, thereby providing more robust constraints on flare–atmosphere interaction physics.

%% Please use the acknowledgment and contribution environments. This will 
%% be anonomyized when the "anonymous" style option is used. 
\begin{acknowledgments}
We are grateful to the anonymous referee and to the editor, Allen W. Shafter, for the careful reading of and useful comments on the manuscript. Yifan Hu is deeply grateful to Prof. Giovanna Tinetti for her valuable support and the insightful suggestions that significantly improved the discussion section. This work made use of data from the NuSTAR mission, a project led by the California Institute of Technology, managed by the Jet Propulsion Laboratory, and funded by the National Aeronautics and Space Administration (NASA). We thank the NuSTAR Operations, Software, and Calibration teams for their support with the execution and analysis of these observations. This research has made use of the NuSTAR Data Analysis Software (NuSTARDAS), jointly developed by the ASI Science Data Center (ASDC, Italy) and the California Institute of Technology (USA). It also utilized data and software provided by the High Energy Astrophysics Science Archive Research Center (HEASARC), a service of the Astrophysics Science Division at NASA’s Goddard Space Flight Center. This work also made use of data provided by NASA through the Swift mission. We thank Dr. James Delaunay, Dr. Kim Page, and the Swift Observatory Duty Scientist for their assistance in coordinating observations with NuSTAR. We also acknowledge the use of data from the Einstein Probe (EP) mission and sincerely thank the EP mission team for supporting our analysis. We are especially grateful to the coordinators and schedulers from Einstein Probe, NuSTAR, and Swift for their generous help in planning and executing the joint observations. We gratefully acknowledge the Principal Investigators of EP, NuSTAR, and Swift for approving our DDT and ToO requests. This project was supported by the Caltech SURF program and NASA grants FAH.NUSBCD-1-NASA.NS. Yuk L. Yung was supported by NASA Grant P2794428 to Caltech.

\end{acknowledgments}

\iffalse
\begin{contribution}
%%This section gives authors the space to recognize author contributions. The text inside this environment is NOT counted towards the total word quanta. At a minimum, manuscripts are expected to include this text:

%% But authors are expected to provide more specific details, e.g. 
%%
%%SC was responsible for writing and submitting the manuscript.
%%WWM came up with the initial research concept and edited the manuscript.
%%OTS obtained the funding and edited the manuscript.
%%EBF provided the formal analysis and validation. He also edited the manuscript.
%%GEH Supervised the undergraduates, wrote the software and administers the project github and Zenodo repositories.
%%
%% Authors can use the Contributor Role Taxonomy (CRediT) at
%% for ideas on how write a good statement tailored to their needs.

\end{contribution}
\fi
\facilities{NuSTAR, Swift-XRT, EP-FXT}

\software{XSPEC \citep{arnaud1996},  
          HEASoft, 
          Python
          }


\bibliography{sample701}{}

\begin{thebibliography}{}
\expandafter\ifx\csname natexlab\endcsname\relax\def\natexlab#1{#1}\fi
\providecommand{\url}[1]{\href{#1}{#1}}
\providecommand{\dodoi}[1]{doi:~\href{http://doi.org/#1}{\nolinkurl{#1}}}
\providecommand{\doeprint}[1]{\href{http://ascl.net/#1}{\nolinkurl{http://ascl.net/#1}}}
\providecommand{\doarXiv}[1]{\href{https://arxiv.org/abs/#1}{\nolinkurl{https://arxiv.org/abs/#1}}}

% type= article
\bibitem[{V.~S. Airapetian {et~al.}(2019)Airapetian, Barnes, Cohen, Collinson, Danchi, Dong, Del~Genio, France, Garcia-Sage, Glocer, Gopalswamy, Grenfell, Gronoff, Güdel, Herbst, Henning, Jackman, Jin, Johnstone, Kaltenegger, Kay, Kobayashi, Kuang, Li, Lynch, Lüftinger, Luhmann, Maehara, Mlynczak, Notsu, Osten, Ramirez, Rugheimer, Scheucher, Schlieder, Shibata, Sousa-Silva, Stamenković, Strangeway, Usmanov, Vergados, Verkhoglyadova, Vidotto, Voytek, Way, Zank, \& Yamashiki}]{Airapetian_2019}
Airapetian, V.~S., Barnes, R., Cohen, O., {et~al.} 2019, \bibinfo{title}{Impact of space weather on climate and habitability of terrestrial-type exoplanets,} International Journal of Astrobiology, 19, 136–194, \dodoi{10.1017/s1473550419000132}

% type= article
\bibitem[{E. Anders \& N. Grevesse(1989)Anders \& Grevesse}]{ANDERS1989197}
Anders, E., \& Grevesse, N. 1989, \bibinfo{title}{Abundances of the elements: Meteoritic and solar,} Geochimica et Cosmochimica Acta, 53, 197, \dodoi{https://doi.org/10.1016/0016-7037(89)90286-X}

% type= inproceedings
\bibitem[{K.~A. Arnaud(1996)Arnaud}]{arnaud1996}
Arnaud, K.~A. 1996, \bibinfo{title}{{XSPEC}: The First Ten Years,} in Astronomical Society of the Pacific Conference Series, Vol. 101, Astronomical Data Analysis Software and Systems V, ed. G.~H. Jacoby \& J.~Barnes, 17

% type= article
\bibitem[{M.~J. Aschwanden {et~al.}(2016)Aschwanden, Holman, O’Flannagain, Caspi, McTiernan, \& Kontar}]{Aschwanden_2016}
Aschwanden, M.~J., Holman, G., O’Flannagain, A., {et~al.} 2016, \bibinfo{title}{GLOBAL ENERGETICS OF SOLAR FLARES. III. NONTHERMAL ENERGIES,} The Astrophysical Journal, 832, 27, \dodoi{10.3847/0004-637x/832/1/27}

% type= article
\bibitem[{A.~O. Benz(2017)Benz}]{Benz2017}
Benz, A.~O. 2017, \bibinfo{title}{Flare Observations,} Living Reviews in Solar Physics, 14, 2, \dodoi{10.1007/s41116-016-0004-3}

% type= article
\bibitem[{P.~T. Boggs {et~al.}(1988)Boggs, Spiegelman, Donaldson, \& Schnabel}]{BOGGS1988169}
Boggs, P.~T., Spiegelman, C.~H., Donaldson, J.~R., \& Schnabel, R.~B. 1988, \bibinfo{title}{A computational examination of orthogonal distance regression,} Journal of Econometrics, 38, 169, \dodoi{https://doi.org/10.1016/0304-4076(88)90032-2}

% type= article
\bibitem[{J.~C. Brown(1971)Brown}]{brown1971deduction}
Brown, J.~C. 1971, \bibinfo{title}{The Deduction of Energy Spectra of Non‐Thermal Electrons in Flares from the Observed Dynamic Spectra of Hard X‐Ray Bursts,} Solar Physics, 18, 489, \dodoi{10.1007/BF00149070}

% type= article
\bibitem[{D.~N. Burrows {et~al.}(2005)Burrows, Hill, Nousek, Kennea, Wells, Osborne, Abbey, Beardmore, Mukerjee, Short, Wood, La~Vallee, Wells, Osborne, \& Abbey}]{burrows2005}
Burrows, D.~N., Hill, J.~E., Nousek, J.~A., {et~al.} 2005, \bibinfo{title}{{The Swift X-Ray Telescope},} Space Science Reviews, 120, 165, \dodoi{10.1007/s11214-005-5097-2}

% type= article
\bibitem[{W. Cash(1979)Cash}]{Cash1979}
Cash, W. 1979, \bibinfo{title}{{Parameter estimation in astronomy through application of the likelihood ratio},} Astrophysical Journal, 228, 939, \dodoi{10.1086/156922}

% type= article
\bibitem[{J. Chadney {et~al.}(2015)Chadney, Galand, Unruh, Koskinen, \& Sanz-Forcada}]{CHADNEY2015357}
Chadney, J., Galand, M., Unruh, Y., Koskinen, T., \& Sanz-Forcada, J. 2015, \bibinfo{title}{XUV-driven mass loss from extrasolar giant planets orbiting active stars,} Icarus, 250, 357, \dodoi{https://doi.org/10.1016/j.icarus.2014.12.012}

% type= article
\bibitem[{E.~W. Cliver {et~al.}(2022)Cliver, Schrijver, Shibata, \& Usoskin}]{Cliver_2022}
Cliver, E.~W., Schrijver, C.~J., Shibata, K., \& Usoskin, I.~G. 2022, \bibinfo{title}{Extreme solar events,} Living Reviews in Solar Physics, 19, \dodoi{10.1007/s41116-022-00033-8}

% type= article
\bibitem[{J.~R.~A. Davenport(2016)Davenport}]{Davenport_2016}
Davenport, J. R.~A. 2016, \bibinfo{title}{THE KEPLER CATALOG OF STELLAR FLARES,} The Astrophysical Journal, 829, 23, \dodoi{10.3847/0004-637x/829/1/23}

% type= misc
\bibitem[{L.~N.~R. do~Amaral {et~al.}(2025)do~Amaral, Shkolnik, Loyd, \& Peacock}]{amaral2025impactstellarflaresatmospheric}
do~Amaral, L. N.~R., Shkolnik, E.~L., Loyd, R. O.~P., \& Peacock, S. 2025, The Impact of Stellar Flares on the Atmospheric Escape of Exoplanets orbiting M stars I: Insights from the AU Mic System, \doarXiv{2503.13353}

% type= manual
\bibitem[{ {EP-FXT Science Data Center}(2025){EP-FXT Science Data Center}}]{epfxt2025}
{EP-FXT Science Data Center}. 2025, {EP-FXT} Users Guide, Version 1.20, Chinese Academy of Sciences.
\newblock \url{http://epfxt.ihep.ac.cn/downloads/FXT_Users_Guide_v1_20-3.pdf}

% type= article
\bibitem[{F. Favata \& G. Micela(2003)Favata \& Micela}]{Favata_2003}
Favata, F., \& Micela, G. 2003, \bibinfo{title}{Stellar Coronal Astronomy,} Space Science Reviews, 108, 577–708, \dodoi{10.1023/b:spac.0000007491.80144.21}

% type= article
\bibitem[{A.~D. Feinstein {et~al.}(2022)Feinstein, France, Youngblood, Duvvuri, Teal, Cauley, Seligman, Gaidos, Kempton, Bean, Diamond-Lowe, Newton, Ginzburg, Plavchan, Gao, \& Schlichting}]{Feinstein_2022}
Feinstein, A.~D., France, K., Youngblood, A., {et~al.} 2022, \bibinfo{title}{AU Microscopii in the Far-UV: Observations in Quiescence, during Flares, and Implications for AU Mic b and c,} The Astronomical Journal, 164, 110, \dodoi{10.3847/1538-3881/ac8107}

% type= article
\bibitem[{D. Foreman-Mackey {et~al.}(2013)Foreman-Mackey, Hogg, Lang, \& Goodman}]{Foreman_Mackey_2013}
Foreman-Mackey, D., Hogg, D.~W., Lang, D., \& Goodman, J. 2013, \bibinfo{title}{<tt>emcee</tt>: The MCMC Hammer,} Publications of the Astronomical Society of the Pacific, 125, 306–312, \dodoi{10.1086/670067}

% type= manual
\bibitem[{K. Forster {et~al.}(2021)Forster, Grefenstette, \& Madsen}]{nustardas2021}
Forster, K., Grefenstette, B., \& Madsen, K. 2021, {NuSTAR} Data Analysis Quickstart Guide, Version 1.2, NuSTAR Science Operations Center, Caltech.
\newblock \url{http://www.srl.caltech.edu/NuSTAR_Public/NuSTAROperationSite/Home.php}

% type= article
\bibitem[{K. France {et~al.}(2013)France, Froning, Linsky, Roberge, Stocke, Tian, Bushinsky, Désert, Mauas, Vieytes, \& Walkowicz}]{France_2013}
France, K., Froning, C.~S., Linsky, J.~L., {et~al.} 2013, \bibinfo{title}{THE ULTRAVIOLET RADIATION ENVIRONMENT AROUND M DWARF EXOPLANET HOST STARS,} The Astrophysical Journal, 763, 149, \dodoi{10.1088/0004-637x/763/2/149}

% type= inproceedings
\bibitem[{N. Gehrels(2004)Gehrels}]{Gehrels_2004}
Gehrels, N. 2004, \bibinfo{title}{The Swift Gamma-Ray Burst Mission,} in AIP Conference Proceedings, Vol. 727 (AIP), 637–641, \dodoi{10.1063/1.1810924}

% type= article
\bibitem[{M. Güdel(2004)Güdel}]{Gudel:2004bz}
Güdel, M. 2004, \bibinfo{title}{X-ray astronomy of stellar coronae,} The Astronomy and Astrophysics Review, 12, \dodoi{10.1007/s00159-004-0023-2}

% type= article
\bibitem[{F.~A. Harrison {et~al.}(2013)Harrison, Craig, Christensen, Hailey, Zhang, Boggs, Stern, Cook, Forster, Giommi, Grefenstette, \& et~al.}]{harrison2013}
Harrison, F.~A., Craig, W.~W., Christensen, F.~E., {et~al.} 2013, \bibinfo{title}{{The Nuclear Spectroscopic Telescope Array (NuSTAR) High‐Energy X‐Ray Mission},} Astrophysical Journal, 770, 103, \dodoi{10.1088/0004-637X/770/2/103}

% type= article
\bibitem[{T.~J. Henry \& W.-C. Jao(2024)Henry \& Jao}]{Henry2024}
Henry, T.~J., \& Jao, W.-C. 2024, \bibinfo{title}{The Character of M Dwarfs,} Annual Review of Astronomy and Astrophysics, 62, 593, \dodoi{10.1146/annurev-astro-052722-102740}

% type= misc
\bibitem[{ {High Energy Astrophysics Science Archive Research Center (HEASARC)}(2025){High Energy Astrophysics Science Archive Research Center (HEASARC)}}]{PIMMS2025}
{High Energy Astrophysics Science Archive Research Center (HEASARC)}. 2025, PIMMS: Portable, Interactive Multi‑Mission Simulator,, \url{https://heasarc.gsfc.nasa.gov/cgi-bin/Tools/w3pimms/w3pimms.pl}

% type= article
\bibitem[{T. Hirayama(1974)Hirayama}]{1974SoPh...34..323H}
Hirayama, T. 1974, \bibinfo{title}{Theoretical Model of Flares and Prominences. I: Evaporating flare model,} Solar Physics, 34, 323, \dodoi{10.1007/BF00153671}

% type= misc
\bibitem[{D.~W. Hogg {et~al.}(2010)Hogg, Bovy, \& Lang}]{hogg2010dataanalysisrecipesfitting}
Hogg, D.~W., Bovy, J., \& Lang, D. 2010, Data analysis recipes: Fitting a model to data, \doarXiv{1008.4686}

% type= article
\bibitem[{G.~D. Holman {et~al.}(2011)Holman, Aschwanden, Aurass, Battaglia, Grigis, Kontar, Liu, Saint-Hilaire, \& Zharkova}]{Holman_2011}
Holman, G.~D., Aschwanden, M.~J., Aurass, H., {et~al.} 2011, \bibinfo{title}{Implications of X-ray Observations for Electron Acceleration and Propagation in Solar Flares,} Space Science Reviews, 159, 107–166, \dodoi{10.1007/s11214-010-9680-9}

% type= article
\bibitem[{S.~R. Kane {et~al.}(2005)Kane, McTiernan, \& Hurley}]{kane2005multispacecraft}
Kane, S.~R., McTiernan, J.~M., \& Hurley, K. 2005, \bibinfo{title}{{Multispacecraft observations of the hard X-ray emission from the giant solar flare on 2003 November 4},} Astronomy and Astrophysics, 433, 1133, \dodoi{10.1051/0004-6361:20041875}

% type= article
\bibitem[{S.~R. Kane {et~al.}(2021)Kane, Foley, Hill, Unterborn, Barclay, Cale, Gilbert, Plavchan, \& Wittrock}]{Kane_2021}
Kane, S.~R., Foley, B.~J., Hill, M.~L., {et~al.} 2021, \bibinfo{title}{Orbital Dynamics and the Evolution of Planetary Habitability in the AU Mic System,} The Astronomical Journal, 163, 20, \dodoi{10.3847/1538-3881/ac366b}

% type= article
\bibitem[{R.~A. Kopp \& G.~W. Pneuman(1976)Kopp \& Pneuman}]{Kopp1976}
Kopp, R.~A., \& Pneuman, G.~W. 1976, \bibinfo{title}{Magnetic reconnection in the corona and the loop prominence phenomenon,} Solar Physics, 50, 85, \dodoi{10.1007/BF00206193}

% type= misc
\bibitem[{A.~F. Kowalski(2024)Kowalski}]{kowalski2024stellarflares}
Kowalski, A.~F. 2024, Stellar flares, \doarXiv{2402.07885}

% type= misc
\bibitem[{A.~F. Kowalski {et~al.}(2024)Kowalski, Osten, Notsu, Tristan, Segura, Maehara, Namekata, \& Inoue}]{kowalski2024risingnearultravioletspectrastellar}
Kowalski, A.~F., Osten, R.~A., Notsu, Y., {et~al.} 2024, Rising Near-Ultraviolet Spectra in Stellar Megaflares, \doarXiv{2411.07913}

% type= article
\bibitem[{M. Kretzschmar(2011)Kretzschmar}]{Kretzschmar_2011}
Kretzschmar, M. 2011, \bibinfo{title}{The Sun as a star: observations of white-light flares,} Astronomy \&amp; Astrophysics, 530, A84, \dodoi{10.1051/0004-6361/201015930}

% type= article
\bibitem[{M. Kretzschmar {et~al.}(2010)Kretzschmar, de~Wit, Schmutz, Mekaoui, Hochedez, \& Dewitte}]{Kretzschmar_2010}
Kretzschmar, M., de~Wit, T.~D., Schmutz, W., {et~al.} 2010, \bibinfo{title}{The effect of flares on total solar irradiance,} Nature Physics, 6, 690–692, \dodoi{10.1038/nphys1741}

% type= article
\bibitem[{H. Lammer {et~al.}(2003)Lammer, Selsis, Ribas, Guinan, Bauer, \& Weiss}]{lammer2003}
Lammer, H., Selsis, F., Ribas, I., {et~al.} 2003, \bibinfo{title}{{Atmospheric Loss of Exoplanets Resulting from Stellar X‐Ray and Extreme‐Ultraviolet Heating},} Astrophysical Journal Letters, 598, L121, \dodoi{10.1086/380815}

% type= article
\bibitem[{U. Mitra-Kraev {et~al.}(2005)Mitra-Kraev, Harra, Güdel, Audard, Branduardi-Raymont, Kay, Mewe, Raassen, \& van Driel-Gesztelyi}]{Mitra_Kraev_2005}
Mitra-Kraev, U., Harra, L.~K., Güdel, M., {et~al.} 2005, \bibinfo{title}{Relationship between X-ray and ultraviolet emission of flares from dMe stars observed by XMM-Newton,} Astronomy \&amp; Astrophysics, 431, 679–686, \dodoi{10.1051/0004-6361:20041201}

% type= article
\bibitem[{R.~A. Murray-Clay {et~al.}(2009)Murray-Clay, Chiang, \& Murray}]{Murray_Clay_2009}
Murray-Clay, R.~A., Chiang, E.~I., \& Murray, N. 2009, \bibinfo{title}{ATMOSPHERIC ESCAPE FROM HOT JUPITERS,} The Astrophysical Journal, 693, 23–42, \dodoi{10.1088/0004-637x/693/1/23}

% type= article
\bibitem[{E.~R. Newton {et~al.}(2016)Newton, Irwin, Charbonneau, Berta-Thompson, Dittmann, \& West}]{Newton_2016}
Newton, E.~R., Irwin, J., Charbonneau, D., {et~al.} 2016, \bibinfo{title}{THE ROTATION AND GALACTIC KINEMATICS OF MID M DWARFS IN THE SOLAR NEIGHBORHOOD,} The Astrophysical Journal, 821, 93, \dodoi{10.3847/0004-637x/821/2/93}

% type= article
\bibitem[{R.~A. Osten {et~al.}(2007)Osten, Drake, Tueller, Cummings, Perri, Moretti, \& Covino}]{Osten_2007}
Osten, R.~A., Drake, S., Tueller, J., {et~al.} 2007, \bibinfo{title}{Nonthermal Hard X‐Ray Emission and Iron Kα Emission from a Superflare on II Pegasi,} The Astrophysical Journal, 654, 1052–1067, \dodoi{10.1086/509252}

% type= article
\bibitem[{R.~A. Osten \& S.~J. Wolk(2015)Osten \& Wolk}]{Osten_2015}
Osten, R.~A., \& Wolk, S.~J. 2015, \bibinfo{title}{CONNECTING FLARES AND TRANSIENT MASS-LOSS EVENTS IN MAGNETICALLY ACTIVE STARS,} The Astrophysical Journal, 809, 79, \dodoi{10.1088/0004-637x/809/1/79}

% type= article
\bibitem[{J.~E. Owen \& A.~P. Jackson(2012)Owen \& Jackson}]{Owen_2012}
Owen, J.~E., \& Jackson, A.~P. 2012, \bibinfo{title}{Planetary evaporation by UV and X-ray radiation: basic hydrodynamics: Planetary evaporation,} Monthly Notices of the Royal Astronomical Society, 425, 2931–2947, \dodoi{10.1111/j.1365-2966.2012.21481.x}

% type= article
\bibitem[{P. Plavchan {et~al.}(2020)Plavchan, Barclay, Gagné, Ciardi, Thao, Crossfield, Mahadevan, Montet, Grieves, Schlieder, Johnson, \& et~al.}]{plavchan2020nature}
Plavchan, P., Barclay, T., Gagné, J., {et~al.} 2020, \bibinfo{title}{{A planet within the debris disk around the pre-main-sequence star AU Microscopii},} Nature, 582, 497, \dodoi{10.1038/s41586-020-2400-z}

% type= article
\bibitem[{J. Poyatos {et~al.}(2025)Poyatos, Fors, Gómez~Cama, \& Ribas}]{Poyatos_2025}
Poyatos, J., Fors, O., Gómez~Cama, J.~M., \& Ribas, I. 2025, \bibinfo{title}{Enhancing the detection of low-energy M dwarf flares: Wavelet-based denoising of CHEOPS data,} Astronomy \&amp; Astrophysics, 699, A242, \dodoi{10.1051/0004-6361/202453517}

% type= article
\bibitem[{F. Reale(2010)Reale}]{Reale_2010}
Reale, F. 2010, \bibinfo{title}{Coronal Loops: Observations and Modeling of Confined Plasma,} Living Reviews in Solar Physics, 7, \dodoi{10.12942/lrsp-2010-5}

% type= article
\bibitem[{J. Sanz-Forcada {et~al.}(2011)Sanz-Forcada, Micela, Ribas, Pollock, Eiroa, Velasco, Solano, \& García-Álvarez}]{Sanz_Forcada_2011}
Sanz-Forcada, J., Micela, G., Ribas, I., {et~al.} 2011, \bibinfo{title}{Estimation of the XUV radiation onto close planets and their evaporation,} Astronomy \&amp; Astrophysics, 532, A6, \dodoi{10.1051/0004-6361/201116594}

% type= article
\bibitem[{J. Sanz‐Forcada {et~al.}(2011)Sanz‐Forcada, Micela, Ribas, Pollock, Eiroa, Velasco, Solano, \& García‐Alvarez}]{sanzforcada2011}
Sanz‐Forcada, J., Micela, G., Ribas, I., {et~al.} 2011, \bibinfo{title}{{Estimating the XUV radiation onto close planets and their evaporation},} Astronomy \& Astrophysics, 532, A6, \dodoi{10.1051/0004-6361/201116594}

% type= article
\bibitem[{C.~J. Schrijver {et~al.}(1995)Schrijver, Mewe, {van den Oord}, \& Kaastra}]{Schrijver1995AandA302}
Schrijver, C.~J., Mewe, R., {van den Oord}, G. H.~J., \& Kaastra, J.~S. 1995, \bibinfo{title}{EUV spectroscopy of cool stars. II. Coronal structure of selected cool stars observed with the EUVE,} Astronomy and Astrophysics, 302, 438

% type= article
\bibitem[{A. Segura {et~al.}(2010)Segura, Walkowicz, Meadows, Kasting, \& Hawley}]{Segura2010}
Segura, A., Walkowicz, L.~M., Meadows, V., Kasting, J., \& Hawley, S. 2010, \bibinfo{title}{The effect of a strong stellar flare on the atmospheric chemistry of an Earth-like planet orbiting an M dwarf,} Astrobiology, 10, 751, \dodoi{10.1089/ast.2009.0376}

% type= article
\bibitem[{K. Shibata \& T. Yokoyama(2002)Shibata \& Yokoyama}]{Shibata_2002}
Shibata, K., \& Yokoyama, T. 2002, \bibinfo{title}{A Hertzsprung‐Russell–like Diagram for Solar/Stellar Flares and Corona: Emission Measure versus Temperature Diagram,} The Astrophysical Journal, 577, 422–432, \dodoi{10.1086/342141}

% type= article
\bibitem[{K. Shibata {et~al.}(2013)Shibata, Isobe, Hillier, Choudhuri, Maehara, Ishii, Shibayama, Notsu, Notsu, Nagao, Honda, \& Nogami}]{Shibata2013PASJ65}
Shibata, K., Isobe, H., Hillier, A., {et~al.} 2013, \bibinfo{title}{Can Superflares Occur on Our Sun?} Publications of the Astronomical Society of Japan, 65, 49, \dodoi{10.1093/pasj/65.3.49}

% type= article
\bibitem[{R.~K. Smith {et~al.}(2001)Smith, Brickhouse, Liedahl, \& Raymond}]{Smith_2001}
Smith, R.~K., Brickhouse, N.~S., Liedahl, D.~A., \& Raymond, J.~C. 2001, \bibinfo{title}{Collisional Plasma Models with APEC/APED: Emission-Line Diagnostics of Hydrogen-like and Helium-like Ions,} The Astrophysical Journal, 556, L91–L95, \dodoi{10.1086/322992}

% type= article
\bibitem[{P.~A. Sturrock(1966)Sturrock}]{Sturrock1966}
Sturrock, P.~A. 1966, \bibinfo{title}{Model of the high-energy phase of solar flares,} Nature, 211, 695, \dodoi{10.1038/211695a0}

% type= article
\bibitem[{P. Testa {et~al.}(2008)Testa, Drake, Ercolano, Reale, Huenemoerder, \& Garc{\'\i}a‐{\'A}lvarez}]{Testa2008}
Testa, P., Drake, J.~J., Ercolano, B., {et~al.} 2008, \bibinfo{title}{Geometry Diagnostics of a Stellar Flare from Fluorescent X‐rays,} The Astrophysical Journal Letters, 675, L97

% type= article
\bibitem[{I.~I. Tristan {et~al.}(2023)Tristan, Notsu, Kowalski, Brown, Wisniewski, Osten, Vrijmoet, White, Carter, Grady, Henry, Hinojosa, Lomax, Neff, Paredes, \& Soutter}]{Tristan_2023}
Tristan, I.~I., Notsu, Y., Kowalski, A.~F., {et~al.} 2023, \bibinfo{title}{A 7 Day Multiwavelength Flare Campaign on AU Mic. I. High-time-resolution Light Curves and the Thermal Empirical Neupert Effect,} The Astrophysical Journal, 951, 33, \dodoi{10.3847/1538-4357/acc94f}

% type= article
\bibitem[{G.~S. Vaiana {et~al.}(1981)Vaiana, Cassinelli, Fabbiano, Giacconi, Golub, Gorenstein, Haisch, Harnden~Jr., Johnson, Linsky, Maxson, Meekins, Rosner, Seward, Topka, Tucker, Van~Speybroeck, \& Zwaan}]{Vaiana1981}
Vaiana, G.~S., Cassinelli, J.~P., Fabbiano, G., {et~al.} 1981, \bibinfo{title}{Results from an extensive Einstein stellar survey,} The Astrophysical Journal, 245, 163, \dodoi{10.1086/158797}

% type= article
\bibitem[{A.~A. West {et~al.}(2008)West, Hawley, Bochanski, Covey, Reid, Dhital, Hilton, \& Masuda}]{West_2008}
West, A.~A., Hawley, S.~L., Bochanski, J.~J., {et~al.} 2008, \bibinfo{title}{CONSTRAINING THE AGE-ACTIVITY RELATION FOR COOL STARS: THE SLOAN DIGITAL SKY SURVEY DATA RELEASE 5 LOW-MASS STAR SPECTROSCOPIC SAMPLE,} The Astronomical Journal, 135, 785–795, \dodoi{10.1088/0004-6256/135/3/785}

% type= article
\bibitem[{J. Wilms {et~al.}(2000)Wilms, Allen, \& McCray}]{Wilms_2000}
Wilms, J., Allen, A., \& McCray, R. 2000, \bibinfo{title}{On the Absorption of X‐Rays in the Interstellar Medium,} The Astrophysical Journal, 542, 914–924, \dodoi{10.1086/317016}

% type= misc
\bibitem[{J.~M. Wittrock {et~al.}(2023)Wittrock, Plavchan, Cale, Barclay, Ludwig, Schwarz, Mekarnia, Triaud, Abe, Suarez, Guillot, Conti, Collins, Waite, Kielkopf, Collins, Dreizler, Mufti, Feliz, Gaidos, Geneser, Horne, Kane, Lowrance, Martioli, Radford, Reefe, Roccatagliata, Shporer, Stassun, Stockdale, Tan, Tanner, \& Vega}]{wittrock2023validatingaumicroscopiid}
Wittrock, J.~M., Plavchan, P., Cale, B.~L., {et~al.} 2023, Validating AU Microscopii d with Transit Timing Variations, \doarXiv{2302.04922}

% type= incollection
\bibitem[{W. Yuan {et~al.}(2022)Yuan, Zhang, Chen, Ling, \& \dots}]{yuan2022}
Yuan, W., Zhang, C., Chen, Y., Ling, Z., \& \dots. 2022, \bibinfo{title}{The Einstein Probe Mission,} in Handbook of X-ray and Gamma-ray Astrophysics, Living Reference Work Entry (Springer), 1--30, \dodoi{10.1007/978-981-16-4544-0_151-1}

\end{thebibliography}
\bibliographystyle{aasjournalv7}

\section{Appendix information}

\begin{table*}[htb]
  \renewcommand{\arraystretch}{1}
  \small
  \centering
  \begin{threeparttable}
    \caption{Best-fit spectral parameters from the \texttt{tbabs + apec + bremss} model}
    \label{tab:bremss}
    \begin{tabular*}{\textwidth}{@{\extracolsep{\fill}} lccc @{}}
      \toprule
      \textbf{Parameters} & \textbf{Unit} & \textbf{Quiescent} & \textbf{Flare 2}\\
      \hline
      $N_\mathrm{H}$\tnote{a} & $10^{20}$\,cm$^{-2}$
        & $3.9$\tnote{i} & $3.9$\tnote{i} \\
      $T$\tnote{b}
        & MK & $17.3^{+2.5}_{-3.2}$ & $21.1^{+3.5}_{-3.0}$  \\
      $Z$\tnote{c}
        &— & $0.27^{+0.45}_{-0.18}$ & $0.19^{+0.19}_{-0.12}$ \\
      ${EM}$\tnote{d}
        &${10^{51}\,\mathrm{cm^{-3}}}$& $6.5^{+2.3}_{-1.7}$& $10.9^{+3.4}_{-2.4}$ \\
      $T_\mathrm{bremss}$\tnote{e}
        & MK & 290 \tnote{j}& 290 \tnote{j} \\
      ${EM}_\mathrm{bremss}$\tnote{f}
      &${10^{49}\,\mathrm{cm^{-3}}}$& $2.6^{+1.5}_{-1.6}$& $6.3^{+3.3}_{-3.3}$ \\
        
    $N_{\rm dof}$\tnote{g}
        &—&285& 299   \\
    $C_{\rm stat}$\tnote{h}
        &—&270.4& 336.2   \\
    Average flux     & $10^{-12}\,\mathrm{erg\,cm^{-2}\,s^{-1}}$  & $0.7_{-0.1}^{+0.1}$    & $1.8_{-0.1}^{+0.1}$       \\
    Corrected ECF     & $\mathrm{erg\,cm^{-2}\,s^{-1}\,count^{-1}}$
    & \begin{tabular}[t]{@{}l@{}}
                    $6.59 \times 10^{-11}$ (FPM-A)\\
                    $7.59 \times 10^{-11}$ (FPM-B) 
                \end{tabular} 
    & \begin{tabular}[t]{@{}l@{}}
                    $6.38 \times 10^{-11}$ (FPM-A)\\
                    $7.33 \times 10^{-11}$ (FPM-B) 
                \end{tabular}

    \\
  
      \bottomrule
    \end{tabular*}

    \vspace{0.5em}
    \begin{tablenotes}[flushleft]
      \footnotesize
      \item[a] $N_{\rm H}$: hydrogen column density.
      \item[b] $T$: plasma temperature.
      \item[c] Abundance relative to solar.
      \item[d] EM: emission measure, a distance of 9.72 pc \citep{plavchan2020nature}.
      \item[e] The bremsstrahlung temperature.
      \item[f] The emission measure of the bremsstrahlung component.
      \item[g] $N_{\rm dof}$: the number of independent spectral bins minus the number of free model parameters.
      \item[h] $C_{\rm stat}$: The Poisson‐based maximum‐likelihood statistic used to evaluate fit quality when count rates are low.
      \item[i] The hydrogen column density was fixed to $3.9\times10^{20}\,$cm$^{-2}$, consistent with the value used in Table \ref{tab:specfit}.
      \item[j] The temperature of the bremss component is fixed at 25 keV (approximately 290 MK) to better constrain the model.
    \end{tablenotes}

  \end{threeparttable}
\end{table*}

\begin{figure*}[ht]
  \centering
  % Left panel
  \begin{subfigure}[t]{0.48\textwidth}
    \centering
    % Replace with your actual file
    \includegraphics[width=\linewidth]{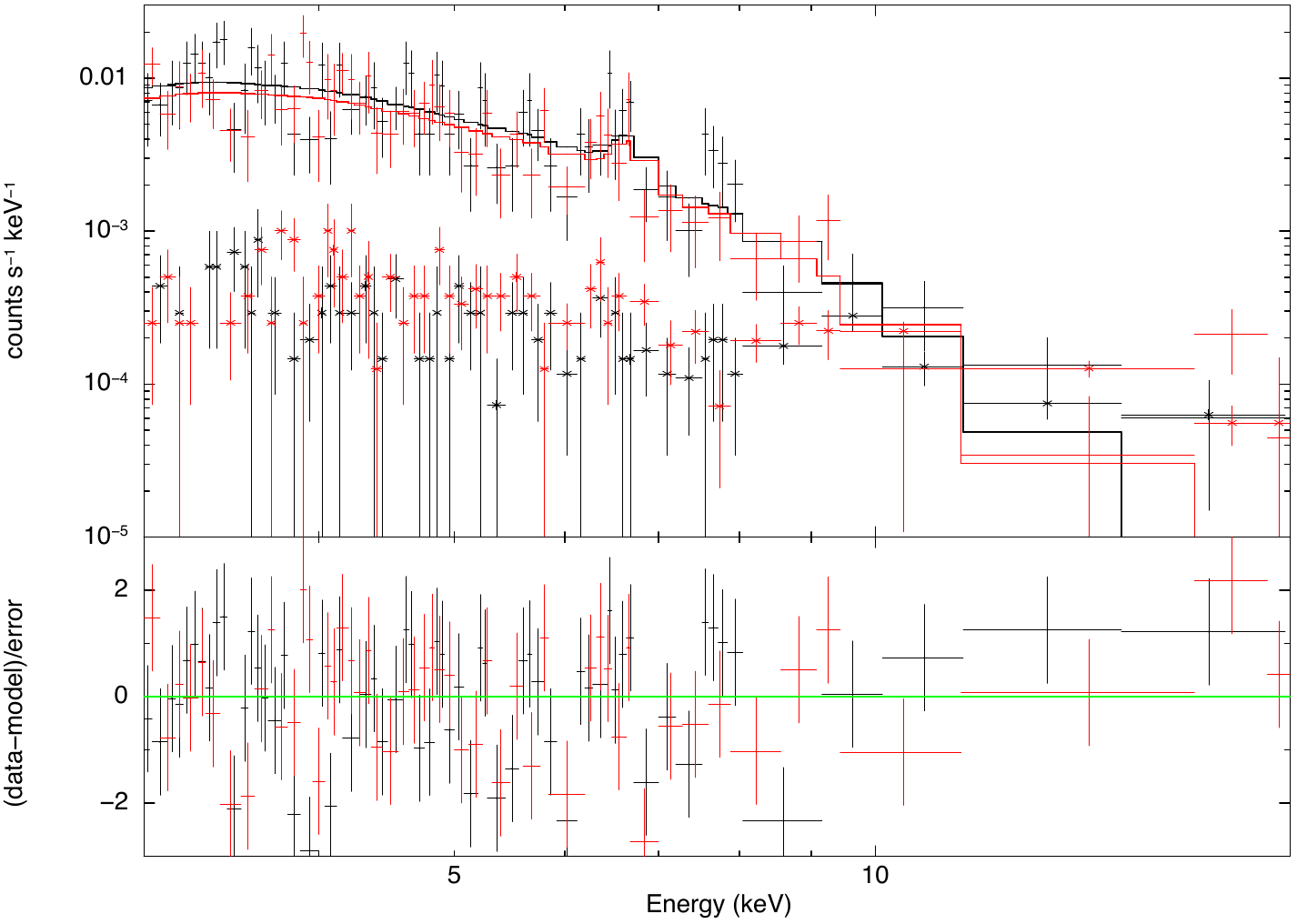}
    \label{fig:flare2_apec_only}
  \end{subfigure}
  \hfill
  % Right panel
  \begin{subfigure}[t]{0.48\textwidth}
    \centering
    % Replace with your actual file
    \includegraphics[width=\linewidth]{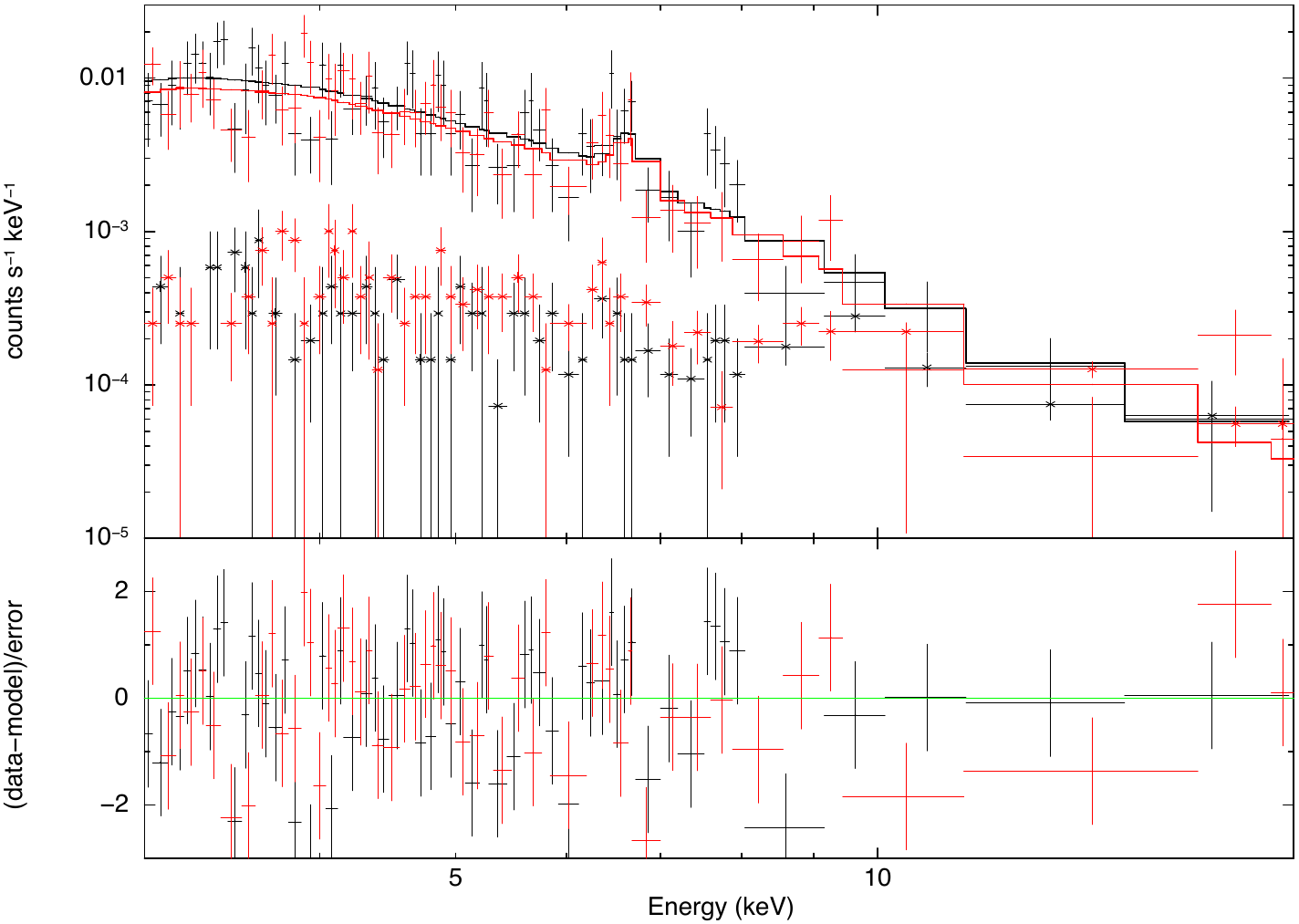}
    \label{fig:flare2_apec_plus_pl}
  \end{subfigure}

  \caption{%
    Flare 2 spectral comparison. Absorbed thermal model \texttt{tbabs+apec} (left) versus absorbed thermal+nonthermal model \texttt{tbabs*(apec+power-law)} (right). In both panels, the observed spectrum (points) is shown together with the best–fit model (solid line) and the background level (star symbols). Adding the power–law reveals a modest high–energy tail relative to a purely thermal fit. After outlier rejection, all data channels above 20 keV lie below the background, indicating that emission beyond this energy is background–dominated. Black points denote FPM–A; red points denote FPM–B.}
  \label{fig:flare2_model_compare}
\end{figure*}



\end{document}